\documentclass[preprint,3p]{elsarticle}

\usepackage{layout}
\usepackage{changepage} 
\usepackage{marvosym}
\usepackage{amssymb}

\usepackage{graphicx}
\usepackage{xcolor}
\usepackage{multicol}
\usepackage{threeparttable}
\usepackage{booktabs}
\usepackage{longtable}
\usepackage{hyperref}
\usepackage{amsthm, amsmath}
\usepackage{algorithm}
\usepackage{algpseudocode}
\usepackage{placeins}
\usepackage{float}
\usepackage{amsfonts}
\usepackage{pgfplots}
\pgfplotsset{compat=1.18} 
\usepackage{subcaption}
\usepackage[normalem]{ulem}
\usepackage[mathlines]{lineno}

\newcommand{\beq}{\begin{equation}}
\newcommand{\eeq}{\end{equation}}

\def\vec   #1{\boldsymbol{#1}}
\newcommand{\smallmath}[1]{\ensuremath{{\scriptstyle #1}}}

\makeatletter
\def\ps@pprintTitle{%
  \let\@oddhead\@empty
  \let\@evenhead\@empty
  \let\@oddfoot\@empty
  \let\@evenfoot\@empty}
\makeatother

\usepackage{bbm}
\usepackage{bm}
\usepackage{mathtools}

\newcommand{\eq}{Eq.}
\newcommand{\eqs}{Eqs.}

\newcommand{\fig}{Fig.}
\newcommand{\figs}{Figs.}

\newcommand{\sect}{Sec.}

\newcommand{\tab}{Tab.}

\renewcommand{\vec}[1]{\bm{#1}} %
\newcommand{\tenII}[1]{\bm{#1}} %
\newcommand{\Space}[1]{\mathbbm{#1}} %

\newcommand{\I}{\tenII{I}} %

\newcommand{\brac}[1]{\left[#1\right]}  %
\newcommand{\mac}[1]{\left<#1\right>}  %
\newcommand{\fOf}[1]{\left(#1\right)}  %
\newcommand{\set}[1]{\left\{#1\right\}}  %

\newcommand{\tr}[1]{\text{tr}\fOf{#1}}  %
\renewcommand{\det}[1]{\text{det}\fOf{#1}}  %
\newcommand{\adj}[1]{\text{adj}\fOf{#1}}  %

\newcommand{\abs}[1]{\left|#1\right|}  %
\newcommand{\norm}[1]{\left|\left|#1\right|\right|}  %

\newcommand{\pd}[2]{\frac{\partial #1}{\partial #2}}  %

\renewcommand{\exp}[1]{\text{exp}\fOf{#1}}

\newcommand{\plaind}{\text{d}}

\newcommand{\TT}{\Space{T}}

\newcommand{\RR}{\Space{R}}

\newcommand{\matVal}[1]{\uppercase{#1}} %

\newcommand{\matVec}[1]{\vec{\matVal{#1}}} %
\newcommand{\vol}[1]{\tilde{#1}}
\newcommand{\iso}[1]{\bar{#1}}

\newcommand{\motion}{\bm{\varphi}}
\newcommand{\F}{\tenII{F}}

\newcommand{\nsv}{n_{\text{s}}}

\newcommand{\StrucVec}{\matVec{a}}

\newcommand{\StrucVeci}{\StrucVec^{i}}
\newcommand{\StrucVecj}{\StrucVec^{j}}

\renewcommand{\u}{\bm{u}}

\renewcommand{\t}{\bm{t}}

\newcommand{\x}{\bm{x}}

\newcommand{\A}{\tenII{A}}

\newcommand{\C}{\tenII{C}}

\renewcommand{\P}{\tenII{P}}

\newcommand{\R}{\tenII{R}}

\newcommand{\X}{\vec{\matVal{x}}}

\renewcommand{\r}{\bm{r}}
\newcommand{\vVec}{\bm{v}}

\newcommand{\en}{\Psi}

\newcommand{\CIso}{\iso{\C}}

\newcommand{\FIso}{\iso{\F}}
\newcommand{\FVol}{\vol{\F}}

\newcommand{\iI}{I_1}
\newcommand{\iII}{I_2}
\newcommand{\iIII}{I_3}
\newcommand{\iIV}{I_4}
\newcommand{\iV}{I_5}

\newcommand{\iIVind}{I_{4,ij}}
\newcommand{\iVind}{I_{5,ij}}

\newcommand{\iIi}{\iso{\iI}}
\newcommand{\iIIi}{\iso{\iII}}

\newcommand{\iIVi}{\iso{\iIV}}
\newcommand{\iVi}{\iso{\iV}}

\newcommand{\iIVIsoind}{\iso{I}_{4,ij}}
\newcommand{\iVIsoind}{\iso{I}_{5,ij}}

\newcommand{\IiIso}{\iso{I}_i}

\newcommand{\dom}{\Omega}
\newcommand{\domO}{\dom_{0}}

\newcommand{\boundN}{\Gamma_N}

\newcommand{\intOn}[2]{\int_{#2}#1\, \plaind #2}

\newcommand{\intOnDomO}[1]{\int_{\domO}#1\, \plaind \domO}

\newcommand{\intOnBoundN}[1]{\intOn{#1}{\boundN}}

\newcommand{\Grad}[1]{\nabla#1}

\newcommand{\act}[1]{\mathcal{#1}}

\newcommand{\uSet}{\act{U}}
\newcommand{\aSet}{\act{A}}
\newcommand{\rSet}{\act{R}}

\newcommand{\actK}{\act{K}}

\newcommand{\nK}{n_{\actK}}

\newcommand{\params}{\bm{\theta}}

\newcommand{\nParams}{n_{\text{p}}}

\newcommand{\FDSet}{\bm{\mathcal{F}}}

\newcommand{\shapeFn}{N}

\newcommand{\shapeFnI}{\shapeFn^{I}}

\newcommand{\nNodes}{{n_{n}}}

\newcommand{\nSteps}{{n_{t}}}

\newcommand{\nBeta}{{n_{\beta}}}

\newcommand{\inter}{\text{int}}
\newcommand{\ext}{\text{ext}}

\begin{document}
% \linenumbers
\begin{frontmatter}

	\title{CANN--EUCLID: unsupervised constitutive artificial neural network model discovery from full-field data}

	\author[inst1,inst2]{Benjamin Alheit}
	\author[inst2]{Siddhant Kumar}
	\author[inst1]{Mathias Peirlinck}

	\affiliation[inst1]{organization={Department of BioMechanical Engineering, Faculty of Mechanical Engineering, Delft University of Technology},
	}
	\affiliation[inst2]{organization={Department of Material Science Engineering, Faculty of Mechanical Engineering, Delft University of Technology},
	}

	\begin{abstract}
    Constitutive artificial neural networks (CANNs) provide an interpretable route to automated material model discovery, but they have so far been used predominantly in stress-supervised settings based on apparent stress--strain data from nominally homogeneous mechanical tests.
    Because each such test samples only a narrow loading path and provides homogenized rather than local stress information, robust constitutive discovery typically requires multiple complementary loading modes to constrain the multidimensional material response.
    This is particularly challenging for soft biological tissues, where repeated testing, damage, and sample-to-sample variability limit the amount of reliable mechanical information that can be extracted from a single specimen.
    Here, we combine CANNs with the stress-unsupervised full-field discovery framework EUCLID to identify sparse hyperelastic constitutive laws directly from displacement fields and reaction forces in a single heterogeneity-inducing loading case.
    The resulting CANN--EUCLID framework minimizes equilibrium imbalance while using sparsity-promoting regularization to select a compact set of active constitutive terms, without requiring local stress measurements or prescribing a specific constitutive law.
    We evaluate our approach on isotropic and anisotropic benchmark problems with prescribed ground-truth laws.
    When the ground-truth law is within the chosen CANN function space, the method recovers the correct active terms with near-exact accuracy, including exponential terms with parameters embedded inside nonlinear functions.
   When it is not contained in the basis, the method retains shared terms and approximates missing contributions using available functions.
    Critically, generalization depends strongly on the deformation states sampled during discovery.
    Exponential strain-stiffening terms can be recovered accurately when sufficiently probed, but can lead to large extrapolation errors when the stiffening regime lies outside the deformation domain of the training data.
    Finally, through comparison of independent forward finite element validation simulations, we show that the behavior discovered by the CANN--EUCLID framework accurately replicates that of the ground truth.
    These results establish stress-unsupervised CANN discovery as a promising framework for interpretable, full-field constitutive model identification.

	\end{abstract}


\end{frontmatter}

\section{Introduction}
\label{sec:intro}

Data-driven constitutive modeling has emerged as a powerful alternative to classical phenomenological constitutive model development in solid mechanics.
Rather than prescribing a constitutive ansatz \textit{a priori} and calibrating a corresponding set of parameters, machine-learning-based constitutive models aim to infer material behavior directly from mechanical testing data while preserving, to varying degrees, the physical structure required for robust prediction and finite element implementation \cite{Ghaboussi_576d,Masi2021_3239,Vlassis2021_7f8b,Tac2022_a940,Tac2022_7978,Linden2023_590d,Dornheim2024_5d67,Fuhg2024_7694,Tac2024_5fa1,Tac2025_1366,Alheit2026_32d4,Jadoon2025_2f3f,Flaschel2026_762c,Martonova2026_20bb,Flaschel2026_3b56}.
Within this broader landscape, hyperelasticity has proven to be a particularly fertile setting for constitutive machine learning because the existence of a strain-energy density provides a natural route to enforce thermodynamic consistency and to derive stresses by differentiation \cite{Linden2023_590d,Klein2022_3243,Magana2025_291c}.
However, the central challenge remains the same: the model must be expressive enough to capture complex nonlinear behavior, while remaining interpretable enough to support insight and generalization.

Consequently, several strategies have been proposed to obtain compact and interpretable hyperelastic models from data.
These approache range from prescribed term or model libraries to symbolic-expression generation and neural network architectures.
Sparse-regression methods prescribe a library of candidate constitutive terms and select a small number of active contributions from that library \cite{Flaschel2023_773f,Urrea_Quintero2026_474e}.
Related candidate-set approaches include Material Fingerprinting, where databases generated from prescribed constitutive model classes and loading protocols are used to identify the material response that best matches experimental data \cite{Flaschel2026_762c,Martonova2026_20bb,Flaschel2026_3b56}.
Symbolic-regression methods represent constitutive laws as symbolic trees of constants, variables, and operators which are mutated by a genetic optimization algorithm \cite{Schoenauer1996_725e,Wang2022_49fb,Abdusalamov2023_6244,Hou2024_1ba4}, whereas grammar-based methods define general rules for generating physically admissible constitutive expressions \cite{Kusner2017_7f94,Kissas2024_1676}.
In parallel, neural constitutive models (NCMs) have increasingly incorporated sparsity and interpretability through architecture design, pruning, or symbolic compression.
Kolmogorov--Arnold-Network (KAN)-based approaches replace fixed activation functions with learnable spline-based functions, which can be inspected or approximated by symbolic expressions after training \cite{Liu2025}.
Such architectures have recently been adapted to constitutive modeling through input-convex KANs \cite{Thakolkaran2025_30c9} and constitutive KAN formulations \cite{Abdolazizi2025}.
Other work has focused on input-convex neural networks (ICNNs), i.e., feed-forward architectures constructed so that the scalar strain energy output remains convex with respect to its inputs. 
These ICNN-based models have been made more compact through smoothed $L_0$ regularization \cite{Fuhg2024_55bc,Anantha_Padmanabha2024_35d1,Tan2026_13fd}.
Related approaches combine pruning with symbolic regression to recover compact analytic expressions from dense neural constitutive models \cite{Bahmani2024_4956,Phan2025_5f41,Sun2025_13f9}.
Taken together, these methods address the same central tension: discovered constitutive laws should be compact and interpretable, but not so restrictive that they fail to represent nonlinear and anisotropic soft-tissue behavior.

Constitutive Artifical Neural Networks (CANNs) provide one attractive compromise in this landscape \cite{Linka2021_1735,Linka2023_6533}.
Rather than selecting among complete pre-specified constitutive laws, CANNs define an expert-structured constitutive model class by combining invariant-based inputs with prescribed nonlinear functions in a branching architecture.
Sparsity-promoting regularization can then suppress inactive branches, yielding strain-energy functions that remain interpretable at the level of individual constitutive terms.
In this sense, CANNs retain the term-level readability of candidate-library approaches, while also allowing gradient-based training and nonlinear-in-parameter representations, including exponential terms that are central to many soft-tissue constitutive laws.
These features make CANNs particularly useful for automated constitutive discovery, where predictive accuracy, physical structure, and mechanistic interpretability must be balanced.
Driven by these advantages, CANNs have seen rapid adoption in biomechanics and soft-tissue modeling, including application to brain \cite{Linka2023_1ca2,Peirlinck2024_f38b,St_Pierre2023}, skin \cite{Linka2023_1bb2}, artificial and real meat \cite{St_Pierre2023_1c60}, arteries \cite{Peirlinck2024_34b9,Vervenne2025_31ed}, and cardiac tissue \cite{Martonova2024_9709,Peirlinck2025_53e6}.
They have also been extended to path-dependent behavior such as viscoelasticity \cite{Wang2023_4e9b,Abdolazizi2024_4363}, growth and remodeling \cite{Holthusen2025_142a}, and Bayesian discovery \cite{Linka2025_32dc}.
Their growing practical relevance is further reflected by integration into commercial finite element workflows through universal Abaqus subroutines \cite{Peirlinck2024_f38b,Peirlinck2024_34b9,Peirlinck2024_55e5}, positioning CANNs as a promising paradigm for interpretable constitutive machine learning.

Despite this progress, CANNs have so far predominantly been trained in a \textit{stress-supervised} setting, in which the network is trained using apparent stress–-strain pairs obtained from mechanical tests such as uniaxial tension, biaxial tension, and triaxial shear \cite{Linka2023_1ca2,Linka2023_1bb2,Martonova2024_9709}.
In this learning paradigm, the experiments are either designed to approximate homogeneous deformation states or interpreted using homogenized force--displacement measures, yielding representative stress--strain data rather than directly measured local stress fields.
While combining multiple loading paths improves identifiability and reduces the risk of overfitting to a single experiment, each test samples only a narrow subset of the admissible deformation space.
As a result, stress-supervised calibration provides only sparse coverage of the multidimensional stress-–strain response of the underlying material.
This sparsity is problematic for constitutive discovery: 
a model may appear accurate along the measured loading paths while remaining weakly constrained elsewhere in the relevant deformation domain.
This limitation is particularly pronounced for soft biological tissues, whose behavior is highly nonlinear, often anisotropic, 
and sensitive to the range and diversity of deformation states used during calibration \cite{Holzapfel2000_7a2d,Gasser2006_3595,Holzapfel2009_4559,Avazmohammadi2019,Martonova2024_9709}.
In practice, obtaining dense local stress data over a sufficiently rich multidimensional deformation space is infeasible for two distinct reasons.
Mechanically, even advanced multiaxial tests probe only restricted deformation manifolds, and stress fields are generally not directly measurable in heterogeneous specimens.
Biologically, the missing deformation-space coverage cannot simply be recovered by repeatedly loading the same specimen, which may induce preconditioning, stress softening, or damage, nor by pooling nominally matched specimens, which is confounded by inter-sample and intra-sample microstructural variability \citep{Aggarwal2023,Krijnen2026_52b3,Famaey2026_7035}.
In contrast, full-field deformation measurements are increasingly accessible through techniques such as digital image correlation and magnetic resonance or ultrasound-based strain imaging \cite{Pierron2012_54c0,Kolawole2023,Meng2025_1d5e,Navy2025_7f0d}, motivating an alternative approach in which constitutive laws are identified directly from full-field kinematics and global equilibrium rather than stress–strain pairs.

In solid mechanics, this idea underlies a long line of stress-unsupervised inverse-identification methods based on full-field measurements.
Classical examples include finite element model updating (FEMU), which uses iterative solutions of costly finite element simulations to update material parameters \cite{Kumar2025_5768,Tan2026_13fd,Knipper2026_560d}; the virtual fields method (VFM), which identifies constitutive parameters through the principle of virtual work \cite{grediac1989principe,Grediac2006_592b,Pierron2012_54c0,Avril2026_1028}; 
and the equilibrium gap method (EGM), which leverages violations of equilibrium as a basis for identification \cite{Claire2004_7994,Amiot2013_290b}.
More recently, Efficient Unsupervised Constitutive Law Identification and Discovery (EUCLID) was introduced to identify sparse interpretable constitutive models by minimizing momentum imbalance while selecting a small number of active terms from an \textit{a priori} library through sparsity-inducing $L_p$ regularization \cite{Flaschel2021_4e47}.
However, its original formulation represents the constitutive response as a linear combination of library terms, which limits its ability to identify models in which material parameters enter nonlinear functions.
This limitation is especially relevant for soft biological tissues, where exponential-type strain-energy functions are ubiquitous \cite{Demiray1972,Humphrey1987_2c2a,Guccione1991_138b,Holzapfel2000_73bc,Krijnen2026_52b3}.
NN-EUCLID addresses part of this limitation by enabling unsupervised training of input-convex neural-network-based constitutive models \cite{Thakolkaran2022_37f6}. 
This increases expressivity, but the resulting models do not retain the sparse term-level interpretability that makes CANNs attractive for constitutive discovery.
Thus, despite substantial advances in the EUCLID family of methods for hyperelasticity, plasticity \cite{Flaschel2022_7c13,Xu2025_4533}, viscoelasticity \cite{Marino2023_5b43}, generalized standard materials \cite{Flaschel2023_564d}, and Bayesian identification \cite{Joshi2022_335f}, a stress-unsupervised EUCLID framework for discovering sparse, interpretable constitutive models with nonlinear parameter dependence remains lacking.

Here, we address this gap by combining CANNs \cite{Linka2023_6533} with the stress-unsupervised discovery framework of NN-EUCLID \cite{Thakolkaran2022_37f6} and the sparsity-promoting $L_p$ regularization strategy of EUCLID \cite{Flaschel2021_4e47}.
The resulting framework discovers CANN-based hyperelastic laws directly from displacement fields and reaction forces, without requiring local stress measurements or a predefined constitutive ansatz.
It thereby brings together the best of all worlds: stress-unsupervised full-field training, sparse term-level interpretability, and nonlinear-in-parameter expressivity.

The remainder of this article is structured as follows.
We present the overall framework for stress-unsupervised training of CANNs in \sect~\ref{sec:methods}.
In \sect~\ref{sec:benchmark}, we assess the framework using synthetic full-field data generated from finite element simulations with prescribed ground-truth constitutive models for isotropic and anisotropic hyperelasticity.
These benchmarks test whether the method can recover the correct constitutive terms when the ground-truth law is contained in the CANN function space, how it approximates ground-truth behavior when required terms are absent from the function space, and under which conditions the discovered models generalize to unseen deformation states.
In \sect~\ref{sec:discussion}, we discuss the implications of the results particularly in the context of soft biological tissues.
Finally, we provide concluding remarks in \sect~\ref{sec:conclusions}.

\section{Methodology}
\label{sec:methods}

\subsection{CANNs}
\label{sec:canns}

The architecture of CANNs is presented in detail in its originating work \cite{Linka2021_1735} with architectural variations proposed in following works
\cite{Linka2023_6533,Vervenne2025_31ed,Martonova2024_9709,Martonova2025_66e5,Peirlinck2024_34b9,Peirlinck2024_f38b,St_Pierre2023,Linka2023_1ca2,Linka2023_1bb2,St_Pierre2023_1c60,Linka2025_32dc,Martonova2026_7570}. 
Here, without loss of generality, we provide a brief description of a particular class of CANN used in this work.
We note that the contributions in this work are not limited to this specific architecture and can be similarly generalized to other CANN architectures.
Details regarding intrinsic enforcement of objectivity, material symmetry, thermodynamic consistency, polyconvexity, etc. are available in the literature \cite{Linka2023_6533,Linden2023_590d,Klein2022_3243,Thakolkaran2025_30c9,Thakolkaran2022_37f6}.
We note that the architecture of CANNs is designed so that these principles are guaranteed in the discovered behavior \citep{Linka2023_6533}.

\subsubsection{Kinematics}

We define the motion of a body $\motion$ as a unique map from points $\X \in \domO \subset \RR^{3}$ to $\x \in \dom \subset \RR^{3}$ at some time $t$, i.e. $\x =\motion\fOf{\X, t}$, where $\domO$ is the initial undeformed domain and $\dom$ is the current deformed domain.
We define the deformation gradient $\F$ and right Cauchy-Green tensor $\C$, respectively, as follows:
\begin{align}
    \F :&= \pd{\motion}{\X}\,, & \C :&= \F ^{T} \F\,.
\end{align}
To satisfy objectivity and material symmetry in the case of hyperelasticity, a strain-energy density function is typically defined in terms of invariants of $\C$.
A typical choice of invariants in the case of anisotropy \cite{Holzapfel2000_73bc} is 
\begin{align}
    \iI & = \tr{\C}\,, & \iII    & = \frac{1}{2}\brac{\tr{\C}^{2} - \tr{\C^{2}}}\,,  & \iIII   & = \det{\C}\,, & \iIVind & = \StrucVeci\cdot \C \StrucVecj\,, & \iVind  & = \StrucVeci\cdot \C^{2}\StrucVecj\,,
\end{align}
where $\StrucVeci$, $i=1,...,\nsv$ is the $i^{\text{th}}$ unit structural vector in the reference configuration, which represents the direction of a microstructural anisotropic constituent, e.g. the axis of a collagen fiber family or the normal of packed myofiber sheets\footnote{When there is only one structural vector, $\nsv=1$, we write $\iIV$ as opposed to $\iIVind$ for the sake of brevity.}.
To capture the significantly different isochoric (volume-preserving) and volumetric responses commonly observed in hyperelastic materials, we multiplicatively decompose the deformation gradient into an isochoric $\FIso$ and volumetric part $\FVol$, i.e.,
\begin{equation}
	\F=\FIso\FVol\,,\qquad \FIso := J^{-1/3}\F\,,\qquad \FVol := J^{1/3}\I\,,\qquad J = \det{\F}\,,
\end{equation}
where $J$ is the (volumetric) Jacobian and $\iso{\bullet}$ and $\vol{\bullet}$ denote the isochoric and volumetric parts of $\bullet$, respectively.
The isochoric right Cauchy-Green tensor is then defined as
\begin{equation}
    \CIso := \FIso ^{T} \FIso\,,
\end{equation}
and the associated isochoric invariants are given by
\begin{align}\label{eq:invs}
    \iIi & = \tr{\CIso}\,, & \iIIi    & = \frac{1}{2}\brac{\tr{\CIso}^{2} - \tr{\CIso^{2}}}\,,  & \iIVIsoind & = \StrucVeci\cdot \CIso \StrucVecj\,, & \iVIsoind  & = \StrucVeci\cdot \CIso^{2}\StrucVecj\,.
\end{align}

\subsubsection{Pseudo-invariants and their requirements}
\label{sec:pseudo-invariants}
Here, we briefly discuss the inputs to CANNs as this aids in their definition that follows in \sect~\ref{sec:cann-architecture}.
The (isochoric) deformation invariants from \eq~\eqref{eq:invs} are not directly input into a CANN, but rather, a set of $n_\actK$ pseudo-invariants $\actK_i$, $i=1,\hdots,n_\actK$. 
The specific form of these pseudo-invariants must be chosen carefully to satisfy the following requirements:

\textit{(i) Zero strain-energy density at $\F=\I$}.
\begin{adjustwidth}{12pt}{}
CANNs are designed such that they evaluate to 0 when all of the inputs are 0.
Consequently, to obtain a material model that results in $\en\fOf{\I}=0$, it is required that $\actK_i\big|_{\F=\I}=0$, $i=1,\hdots,\nK$. \newline
\end{adjustwidth}

\textit{(ii) Zero stress at $\F=\I$}.
\begin{adjustwidth}{12pt}{}
The first Piola-Kirchhoff stress is given by 
\begin{equation}
    \P = \pd{\en}{\F} = \sum_i^{\nK}\pd{\en}{\actK_i}\pd{\actK_i}{\F}\,,
\end{equation}
which, in the special case that a given $\actK_i$ is defined solely in terms of isochoric invariant $\IiIso$, expands to
\begin{equation}
    \P =  \sum_i^{\nK}\pd{\en}{\actK_i}\pd{\actK_i}{\IiIso}\pd{\IiIso}{\F}\,.
\end{equation}
To ensure that $\P\fOf{\I}=\bm{0}$, it is required that at least one of the factors $\pd{\en}{\actK_i}$, $\pd{\actK_i}{\IiIso}$, or $\pd{\IiIso}{\F}$ evaluate to $0$ when $\F=\I$.
In general, CANNs do not ensure that $\pd{\en}{\actK_i}\big|_{\F=\I}=0$, hence when using isochoric invariants either $\pd{\actK_i}{\IiIso}$, or $\pd{\IiIso}{\F}$ must be zero at $\F=\I$. 
For $\iIi$ and $\iIIi$ this is trivial as $\pd{\iIi}{\F}\big|_{\F=\I}=\pd{\iIIi}{\F}\big|_{\F=\I}=\bm{0}$. 
However, this is not the case for $\iIVIsoind$ and $\iVIsoind$.
Similarly, for pseudo-invariants $\actK_i$ that reflect $J$, we have $\pd{J}{\F}\big|_{\F=\I}\neq\bm{0}$.
In these cases, the expression for the pseudo-invariants $\actK_i$ must be chosen such that $\pd{\actK_i}{\IiIso}\big|_{\F=\I}=0$.
\end{adjustwidth}

\textit{(iii) Polyconvexity}.
\begin{adjustwidth}{12pt}{}
    Polyconvexity of the strain-energy function \cite{Ball1976_4d15} is the \textit{de facto} sufficient condition used to guarantee the existence of a solution to a finite strain elasticity boundary value problem.
    Hence, the CANN should ultimately result in a polyconvex strain-energy function. 
    As is detailed in the following subsection (\sect~\ref{sec:cann-architecture}), the CANN architecture generates a sum of interpretable terms by composing various activation functions that are applied to the pseudo-invariant inputs.
    To ensure that the overall discovered behavior is polyconvex, one must ensure that each resulting term in the sum is polyconvex.
    As such, the pseudo-invariants must be chosen appropriately. 
    We show that the choices for the activation functions and pseudo-invariants used in this work result in a polyconvex strain-energy density function in \ref{sec:cann-polyconvexity}.

\end{adjustwidth}
In this work, without loss of generality, we choose the following pseudo-invariants as inputs to the CANN:
\begin{align}
    \actK_1 &= \iIi -3\,,
            & \actK_2 &= \iIIi^{3/2} - 3 ^{3/2}\,,
            & \actK_3 &= \brac{J-1}^{2}\,,
            & \actK_{4,ij} &= \begin{cases}
                \mac{\iIVIsoind - 1}^{2} & \text{if } i=j\,,\\ 
                \iIVIsoind^{2} & \text{otherwise}\,.
            \end{cases} 
            \label{eq:used-psuedo-invariants}
\end{align}
However, other choices are also possible, see e.g. \cite{St_Pierre2023,Martonova2025}. 
Moreover, for the sake of polyconvexity, we use $\iIIi^{3/2}$ as opposed to $\iIIi$ \cite{Schroder2003_1753} and $\mac{\iIVIsoind - 1}^{2}$ as opposed to $\brac{\iIVIsoind - 1}^{2}$ \cite{Balzani2006_50ce}, see \ref{sec:cann-polyconvexity}.

\subsubsection{CANNs map invariants to strain-energy density}
\label{sec:cann-architecture}
\begin{figure}[!htb]
	\begin{center}
		\includegraphics[width=0.75\textwidth]{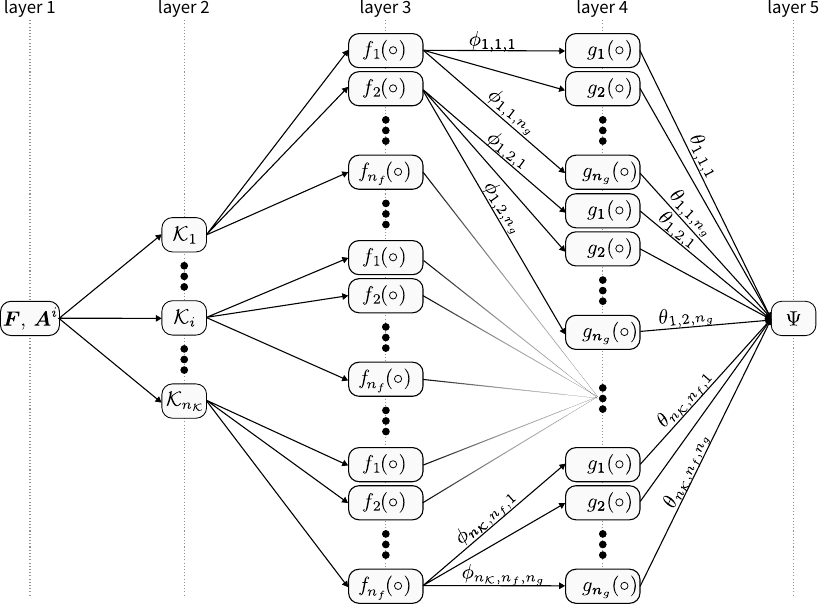}
	\end{center}
    \caption{\textbf{Constitutive Artificial Neural Network (CANN) architecture.} 
    CANNs define a map from the deformation gradient $\F$ and structural vectors $\StrucVeci$ to the strain-energy density $\Psi$.
    This is done by defining a branching tree that automatically generates a large sum of interpretable terms.
    First, a set of $n_\actK$ scalar invariants is obtained as defined in \eq~\eqref{eq:used-psuedo-invariants}.
    These are then operated on by a set of functions $f_{j}$, $j=1,\hdots,n_f$.
    The outputs of these functions are then multiplied by weights $\phi_{ijk}$ and fed into a following set of functions $g_{k}$, $k=1,\hdots,n_g$.
    Finally, the outputs of the functions $g_{k}$ $k=1,\hdots,n_g$ are multiplied by a second set of weights $\theta_{ijk}$ and summed to provide the final strain-energy density.}
	\label{fig:cann-arch}
\end{figure}

CANNs provide a structured neural architecture for representing the strain-energy density as a sum of interpretable terms.
The computational graph for achieving this is illustrated in \fig~\ref{fig:cann-arch}.
In contrast to the densely connected layers typical of many neural network architectures, CANNs utilize a sparse, branching tree-like architecture.
An additional contrast to conventional neural network architectures is that CANNs typically have a fixed depth of five layers, with each layer corresponding to a level of the branching tree.

The first layer of the CANN takes the deformation gradient $\F$ and structural vectors $\StrucVeci$, $i=1,\hdots,\nsv$ as input to the network.
A set of pseudo-invariants, as discussed in \sect~\ref{sec:pseudo-invariants}, is calculated in the second layer, resulting in $\nK$ nodes.
In the third layer, a set of functions $f_{1}\fOf{\circ},\hdots,f_{n_f}\fOf{\circ}$ are applied to each pseudo-invariant from the second layer, leading to $\nK\times n_f$ nodes.
The output of each node in the third layer is replicated $n_g$ times, multiplied by a unique weight $\phi_{ijk}$ ($i\in[1,\nK]$, $j\in[1,n_f]$, $k\in[1,n_g]$), and fed into a corresponding function in the fourth layer $g_{1}\fOf{\circ},\hdots,g_{n_g}\fOf{\circ}$, leading to $\nK\times n_f \times n_g$ nodes.
The output of each node in the fourth layer is multiplied by a unique weight $\theta_{ijk}$ and summed to provide the strain-energy density as output in the fifth layer.
The indices of the weights $\phi_{ijk}$ and $\theta_{ijk}$ associated with selected edges have been indicated in \fig~\ref{fig:cann-arch}\footnote{When referring to specific components of $\phi_{ijk}$ and $\theta_{ijk}$, we use commas between indices to disambiguate cases where one or more indices are larger than nine. 
For example, when writing $\phi_{1213}$ the values of specific indices are unclear ($i=1$, $j=21$, $k=3$, or $i=12$, $j=1$, $k=3$, etc.), so instead we write, e.g., $\phi_{1,21,3}$ which clarifies that $i=1$, $j=21$, and $k=3$.}.

The mathematical description of the process illustrated visually in \fig~\ref{fig:cann-arch} and textually in the previous paragraph is concisely represented by the following equation:
\begin{equation}
    \en\fOf{\F, \StrucVec ^{1},\hdots,\StrucVec ^{\nsv}} := \sum_{i}^{n_{\actK}}\sum_{j} ^{n_f}\sum_{k}^{n_g}\underbrace{\theta_{ijk}g_{k}\fOf{\phi_{ijk}f_{j}\fOf{\actK_i \fOf{\F, \StrucVec ^{1},\hdots,\StrucVec ^{\nsv}}}}}_{\en_{ijk}}\,,
	\label{eq:cann-def}
\end{equation}
which clearly yields a sum of $\nK\times n_f \times n_g$ terms, or \textit{constitutive neurons} $\en_{ijk}$ \cite{Peirlinck2024_55e5}.

To create a concretized instance of a CANN from this abstract description, one must choose the set of pseudo-invariants to use $\actK_i$, $i=1,\hdots,\nK$, and functions $f_{1}\fOf{\circ},\hdots,f_{n_f}\fOf{\circ}$ and $g_{1}\fOf{\circ},\hdots,g_{n_g}\fOf{\circ}$.
In this work, we choose to use the pseudo-invariants listed in \eq~\eqref{eq:used-psuedo-invariants}.
A typical choice for the functions $f_{1}\fOf{\circ},\hdots,f_{n_f}\fOf{\circ}$, which is also used in this work, is
\begin{equation}
    f_{j}\fOf{\circ} = \circ^{j}\,.
	\label{eq:f1}
\end{equation}
For the functions $g_{1}\fOf{\circ},\hdots,g_{n_g}\fOf{\circ}$, we choose 
\begin{align}
    g_{k}\fOf{\circ} & = \begin{cases}
        \circ & \text{if } k=1\,, \\
        \exp{\circ}-1 & \text{if } k=2\,.
	        \end{cases}
	\label{eq:f2}
\end{align}
In ~\ref{sec:cann-polyconvexity}, we show that these specific activation function choices and pseudo-invariants yield a polyconvex strain-energy density, provided that the CANN weights are non-negative. Accordingly, throughout training we constrain all trainable weights $\theta_{ijk}$ and $\phi_{ijk}$ to satisfy $\theta_{ijk},\phi_{ijk}\geq0$.

\subsection{CANNs meet unsupervised discovery}

\subsection*{Minimizing imbalance of equilibrium}
Following prior EUCLID works \cite{Flaschel2021_4e47,Thakolkaran2022_37f6}, as a starting point, we assume that one has a dataset of displacement values measured at a finite number of material points $\nNodes$ that correspond to nodes in a FE mesh, and a finite number of time steps $\nSteps$, i.e.
\begin{equation}
    \uSet = \set{\u^{I,t} \in \RR^{3} \hspace{0.5em}|\hspace{0.5em} I \in \brac{1,\,\nNodes},\, t \in \brac{1,\,\nSteps}}\,,
\end{equation}
and, if the material is anisotropic, we also assume that one has access to an approximation of the structural vector fields which we denote collectively as $\aSet$.
Additionally, we assume that one has access to the aggregated reaction forces on $\nBeta$ boundaries at those time steps, i.e.
\begin{equation}
    \rSet = \set{\R^{\beta,t} \in \RR^{3} \hspace{0.5em}|\hspace{0.5em} \beta \in \brac{1,\,\nBeta},\, t \in \brac{1,\,\nSteps}}\,. 
    \label{eq:reaction-data}
\end{equation}
The goal is to discover the constitutive model and parameters that satisfy the weak form of equilibrium balance, i.e. to find $\params = \set{\fOf{\phi_{ijk},\theta_{ijk}} \hspace{0.5em}|\hspace{0.5em} i\in[1,\nK],\, j\in[1,n_f],\, k\in[1,n_g]}$ that satisfies
\begin{equation}
    \intOnDomO{\P\fOf{\F, \aSet; \params} : \Grad{\vVec}} - \intOnBoundN{\t \cdot \vVec} = 0\,, \qquad \forall \vVec\,,
    \label{eq:weak-form-continuous}
\end{equation}
where $\vVec$ is an arbitrary test function from the same function space as $\u$ and we have assumed that body forces are negligible.
To obtain a continuous field for $\u$ from the discrete measurements in $\uSet$, we use basis functions $\shapeFnI$ associated with each node $I$ that define a piecewise polynomial function space over the domain, i.e.
\begin{equation}
    \u^{t} \approx \sum_I ^{\nNodes} \shapeFnI  \u ^{I,t}\,,
    \label{eq:approx-u}
\end{equation}
where $\bullet^{t}$ denotes the value of a field at time $t$.
This allows for obtaining an approximation for the deformation gradient over the domain as follows
\begin{equation}
    \F^{t} =  \I + \Grad{\u}^{t} \approx \I + \sum_I ^{\nNodes} \u ^{I,t} \otimes \Grad{\shapeFnI}\,.
    \label{eq:f-calc}
\end{equation}
Since $\u$ and $\vVec$ are from the same function space, we choose the same basis functions to provide an approximation for the test function $\vVec$, i.e. 
\begin{equation}
    \vVec \approx \sum_I ^{\nNodes} \shapeFnI  \vVec ^{I}\,.
    \label{eq:approx-v}
\end{equation}
Substituting \eqs~\eqref{eq:approx-u} and \eqref{eq:approx-v} into \eq~\eqref{eq:weak-form-continuous} results in the following discretized version of the weak form of momentum balance:
\begin{equation}
    \sum_{I}^{\nNodes} v^{I}_{i} \brac{\intOnDomO{P\fOf{\F^{t}, \aSet; \params}_{ij} \Grad{\shapeFnI}_{j}} - \intOnBoundN{t^{t}_i \shapeFnI}} = 0\,, \qquad \forall v^{I}_{i}.
    \label{eq:weak-form-disc}
\end{equation}
Due to the arbitrariness of $v^{I}_{i}$, \eq~\eqref{eq:weak-form-disc} holds if and only if the expression in parentheses is zero for all $(I,i)$.
Accordingly, we interpret this as an unbalanced force at each node $r^{I,t}_{i}$, i.e. 
\begin{equation}
    r^{I,t}_i := \underbrace{\intOnDomO{P\fOf{\F^{t}, \aSet; \params}_{ij} \Grad{\shapeFnI}_{j}}}_{r_{\inter,i}^{I,t}} - \underbrace{\intOnBoundN{t^{t}_i \shapeFnI}}_{r_{\ext,i}^{I,t}} = 0\,,
    \label{eq:force-imbalance}
\end{equation}
where we further identify internal and external unbalanced forces denoted by $r_{\inter,i}^{I,t}$ and $r_{\ext,i}^{I,t}$, respectively, for node $I$ in direction $i$ at time $t$.
\eq~\eqref{eq:force-imbalance} provides a basis for an objective function that can be minimized with respect to material parameters $\params$ to approximately satisfy balance of equilibrium, namely
\begin{equation}
    L = \sum_{I,i,t} \brac{r_i^{I,t}}^{2}\,.
    \label{eq:obj-all}
\end{equation}
However, in practice, one does not typically know the traction $\t$ on the constrained surfaces and so $r_{i}^{I,t}$ cannot be determined for the nodes on those surfaces.
To circumvent this issue we note that although one does not have access to the traction, one does have access to the reaction force for the constrained surfaces, i.e. \eq~\eqref{eq:reaction-data}.
Additionally, since the internal and external forces should balance pointwise, their sums should also balance over a given boundary, i.e.
\begin{equation}
    R^{\beta,t}_i = \sum_{I \in D^{\beta}} r_{\ext, i}^{I,t} = \sum_{I \in D^{\beta}} r_{\inter, i}^{I,t}\,,
\end{equation}
where $D^{\beta}$ denotes the set of nodes on boundary $\beta$.
Accordingly, instead of minimizing \eq~\eqref{eq:obj-all}, we additively decompose the objective function into a component for the free and fixed nodes, respectively,
\begin{align}
    L_{\inter} &= \frac{1}{\nSteps\nNodes} \sum_{t}\sum_{(I,i) \in D^{\text{free}}} \brac{r_{\inter, i}^{I,t}} ^{2} \,,\label{eq:l-free}\\
    L_{\ext} &= \frac{1}{\nSteps\nBeta} \sum_{\beta,t} \brac{\R^{\beta,t}-\sum_{I \in D^{\beta}} \r_{\inter} ^{I,t}}^{2} \,,\label{eq:l-fixed}
\end{align}
where $ D^{\text{free}}$ denotes all nodes that are not constrained.

\subsection*{Sparsity-promoting regularization}

The goal of model discovery is not only to identify a model that minimizes the imbalance of equilibrium, but also to select, among admissible fits, a compact constitutive representation with a small number of interpretable and physically meaningful terms. This requires balancing accuracy against sparsity, and thus predictability against interpretability.
The \textit{de facto} approach for achieving this in the context of constitutive modeling is to penalize the number of non-zero terms in a discovered model by adding the $L_p$-norm of the model coefficients to the loss equation, i.e.
for a model of the form 
\begin{equation}
    h\fOf{x} = \sum_i ^{\nParams}\alpha_i h_i \fOf{x},
    \label{eq:example-func}
\end{equation}
the $L_p$-norm of its coefficients is given by
\begin{equation}
    \norm{\bm{\alpha}}_p =  \brac{\sum_{i}^{\nParams} \abs{\alpha_i}^{p}}^{1/p}\,,
\end{equation}
where $\nParams$ is the number of parameters.
In particular, this penalizes the number of non-zero coefficients for $0\leq p\leq1$.
Using such a regularization term to promote sparsity was first suggested in \cite{Frank1993_6411} as a generalization of $L_2$ regularization, i.e. $L_p$ regularization with $p=2$.
The $L_2$ case was originally proposed in \cite{Hoerl1970_614f} to stabilize regression problems with a high degree of collinearity rather than to promote sparsity.
This idea was first applied in the context of constitutive model discovery in the original EUCLID work \cite{Flaschel2021_4e47} and has since been adopted extensively in the training of CANNs \cite{McCulloch2024_6c1e,Martonova2025_66e5,Martonova2024_9709,Linka2023_1ca2}.

We note that the stress given by a CANN (the strain-energy density for which is given in \eq~\eqref{eq:cann-def}) is given by
\begin{equation}
    \P = \pd{\en}{\F} = \sum_{i,j,k} \pd{\en_{ijk}}{\F} = \sum_{i}^{\nK} \sum_{j}^{n_f} \sum_{k}^{n_g} \underbrace{\theta_{ijk}\phi_{ijk}}_{\equiv\alpha_i \text{in \eq~\eqref{eq:example-func}}} \underbrace{g'_{k}f'_{j} \pd{\actK_i}{\F}}_{\equiv h_i \text{in \eq~\eqref{eq:example-func}}}\,, \label{eq:cann-stress}
\end{equation}
where $\bullet'$ denotes the derivative of a function $\bullet$ and we identify that $\theta_{ijk}\phi_{ijk}$ is equivalent to $\alpha_i$ in \eq~\eqref{eq:example-func} and $g'_{k}f'_{j} \pd{\actK_i}{\F}$ is equivalent to $h_i$ in \eq~\eqref{eq:example-func}.
As such, we propose that the appropriate regularization term to use is 
\begin{equation}
    L_{p} = \frac{1}{\nK n_{f}n_{g}}\sum_{i}^{n_{\actK}}\sum_{j} ^{n_f}\sum_{k}^{n_g}\abs{\theta_{ijk}\phi_{ijk}}^{p}\,.
    \label{eq:our-reg}
\end{equation}
Hence, the final loss equation is given by
\begin{equation}
    L = \lambda_{\inter} L_{\inter} +\lambda_{\ext} L_{\ext} +\lambda_{p} L_{p}
    \label{eq:loss}
\end{equation}
where $L_{\inter}$ penalizes the imbalance of equilibrium on unconstrained nodes (\eq~\eqref{eq:l-free}), 
$L_{\ext}$ 
penalizes the imbalance of reaction forces on constrained nodes (\eq~\eqref{eq:l-fixed}),
$L_p$ penalizes the number of non-zero terms in the final model,
and $\lambda_{\inter}$, $\lambda_\ext$, and $\lambda_p$ weight the relative contributions of 
$L_\inter$, $L_\ext$, and $L_p$, respectively, to the overall loss.

\begin{figure}[tb]
	\begin{center}
		\includegraphics[width=\textwidth]{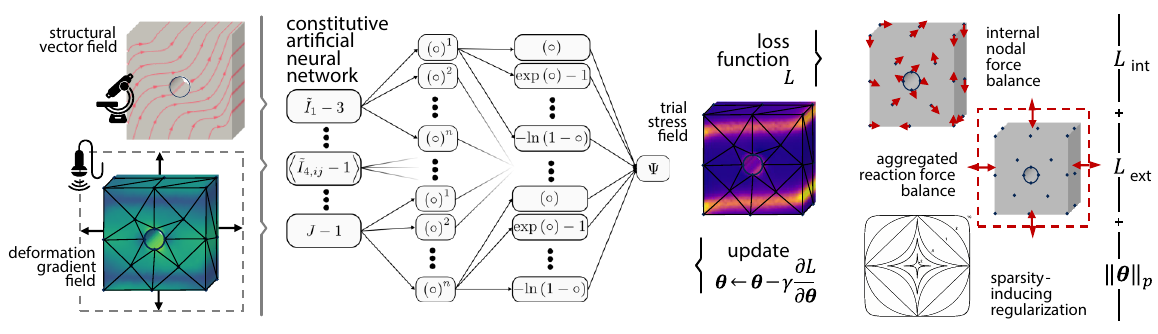}
	\end{center}
    \caption{\textbf{Schematic summary of the CANN--EUCLID framework.}
        The structural vector and deformation gradient fields, which are assumed to be measured data, act as inputs to a CANN. 
        The CANN maps these fields to a strain-energy density, from which a trial stress field is obtained by differentiating the strain-energy density with respect to the input deformation gradient $\F$.
        The imbalance of equilibrium for the trial stress field is calculated and added to the loss, in addition to the imbalance of the reaction forces and the sparsity-promoting $L_p$-norm of the model parameters \cite{Frank1993_6411}.
        If the loss is not below some tolerance, then the weights are updated using a gradient-based optimization algorithm.
        This process repeats until the loss is below the chosen tolerance, at which point we conclude that a sparse CANN has been identified that satisfies the internal balance of equilibrium and reaction forces within an acceptable tolerance.}
	\label{fig:training-diagram}
\end{figure}

With the loss \eq~\eqref{eq:loss} in hand, the overall CANN--EUCLID training procedure is summarized in \fig~\ref{fig:training-diagram}.
The structural vector and deformation gradient fields, which are assumed to be measured data, act as inputs to a CANN. 
The CANN maps these fields to a strain-energy density, from which a trial stress field is obtained by differentiating the strain-energy density with respect to the input deformation gradient $\F$. 
The imbalance of equilibrium for the trial stress field is calculated and added to the loss, in addition to the imbalance of the reaction forces and the sparsity-promoting $L_p$-norm of the model parameters. 
If the loss is not below a chosen tolerance, then the weights are updated using a gradient-based optimization algorithm. 
This process repeats until the loss is below the chosen tolerance, at which point we conclude that a sparse CANN has been identified that satisfies the internal balance of equilibrium and reaction forces within an acceptable tolerance. 

In practice, directly optimizing \eqref{eq:loss} with sparsity-promoting regularization can be sensitive to the initialization of the CANN weights, as well as to the scale of the deformations and reaction forces in the data.
To address these issues, we introduce a \textit{data-normalization strategy} suitable for the stress-unsupervised setting, together with appropriate CANN \textit{weight initialization} and \textit{multistage training} procedures.
These strategies, along with additional training details, are described in Section~\ref{sec:training-details}.
In our experience, these strategies significantly improved the robustness of stress-unsupervised training of the CANN.

\subsection{Evaluation of the CANN--EUCLID framework}

\subsubsection{Synthetic data generation}

\begin{figure}[tb]
    \begin{center}
        \includegraphics[width=.75\textwidth]{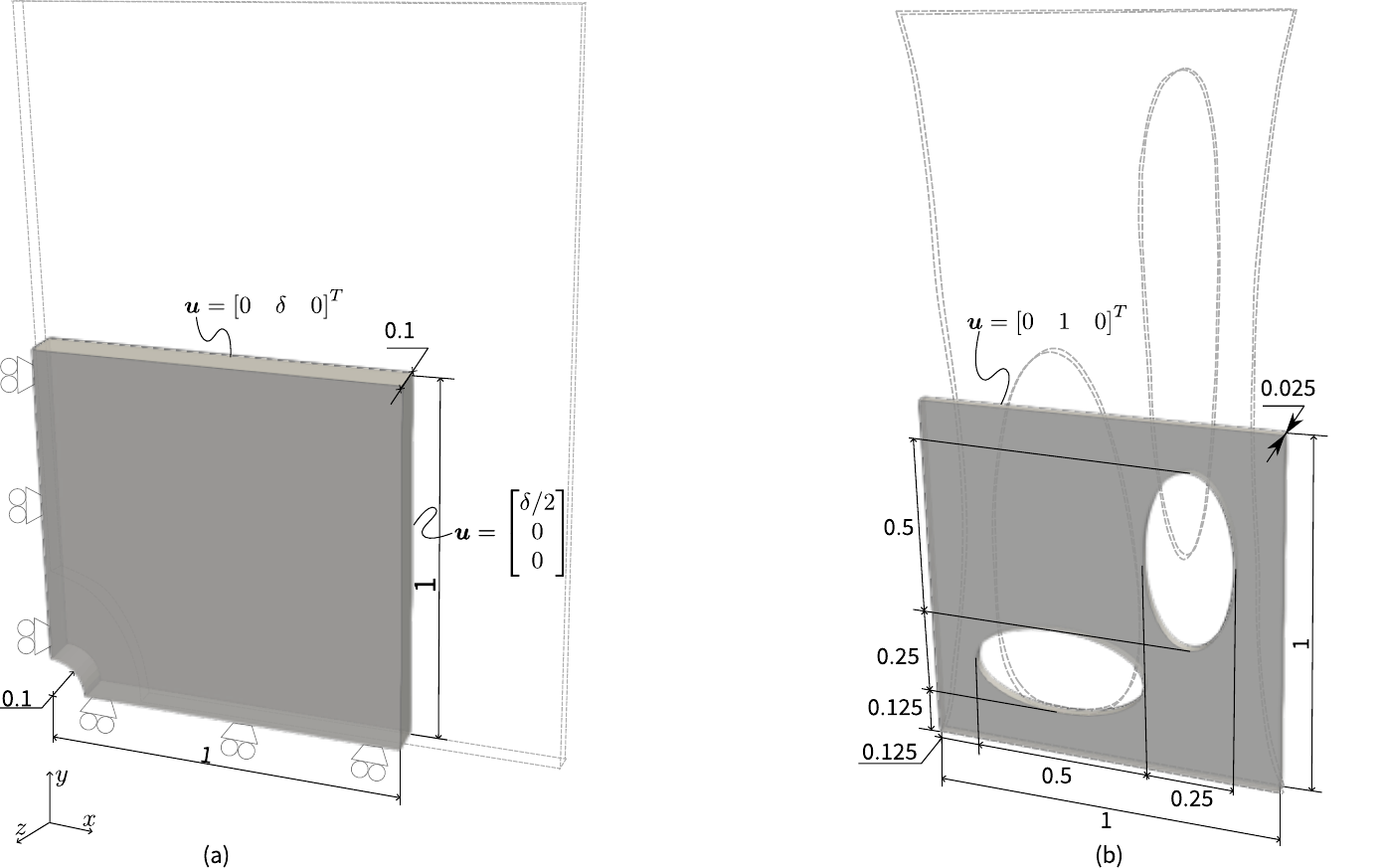}
    \end{center}
    \caption{\textbf{Schematic illustration of geometries used in training and validation.}
        The training geometry (a) consists of a plate with a hole in the bottom left corner.
        It is subjected to vertical displacement on the top face, 
        horizontal displacement on the right face,
        and roller boundary conditions on the left and bottom faces. 
        The amount of deformation is parameterized by $\delta$ which is linearly ramped from zero to one over ten time steps.
        The validation geometry (b) consists of a plate with elliptical holes.
        It is subject to a unit of vertical displacement on the top face while the bottom face remains fixed.
        All other boundaries are traction free.
    }
    \label{fig:geoms}
\end{figure}

We verify the success of the unsupervised training of CANNs using canonical benchmarks for unsupervised constitutive behavior discovery, as established in \cite{Flaschel2021_4e47,Thakolkaran2022_37f6,Thakolkaran2025_30c9}, for both isotropic and anisotropic hyperelasticity.
To this end, we generate synthetic displacement field and reaction force data by applying biaxial loading to a plate with a central hole, as illustrated in \fig~\ref{fig:geoms}~(a).
The amount of deformation is parameterized by $\delta$, which is linearly ramped from zero to one over ten time steps (see \ref{sec:hyperparameters} for details).
These data are obtained from \texttt{COMMET-solve} FE simulations \cite{Alheit2026_32d4} with chosen ground-truth constitutive models.
In the isotropic cases the ground-truth constitutive models considered 
\footnote{We replace the original material parameters by arbitrary benchmark values and write the models directly to avoid introducing notation used only once.}
are as follows
\footnote{Models that contain exponential terms are annotated with ``*'' throughout this article for ease of reference.}:
\begin{subequations}
\begin{align}
&\text{Neohookean \cite{Rivlin1948_7623}: } \nonumber\\
&\quad \en = 0.5 \brac{\iIi-3} + 1.5\brac{J-1} ^{2}\,, \\
&\text{Demiray *\cite{Demiray1972}: } \nonumber\\
&\quad \en = 0.5\brac{\exp{0.5 \brac{\iIi -3}}-1}+ 1.5\brac{J-1} ^{2} \,,\\
&\text{Isihara \cite{Isihara1951_5512}: } \nonumber\\
&\quad \en = 0.5 \brac{\iIi-3} +\brac{\iIIi-3} + \brac{\iIi-3} ^{2} + 1.5\brac{J-1} ^{2}\,,\\
&\text{Arruda-Boyce \cite{Arruda1993_3631}: } \nonumber\\
&\quad \en = 2.5 \sqrt{N_c}\brac{\beta_c \lambda_c - \sqrt{N_c} \log \fOf{\frac{\sinh \beta_c}{\beta_c}}} - c_{AB} + 1.5\brac{J-1} ^{2} \,, \\
&\text{Gent-Thomas \cite{Gent1958_2fcf}: }\nonumber \\
&\quad \en = 0.5 \brac{\iIi-3} + \log\fOf{\iIIi/3} + 1.5\brac{J-1} ^{2} \,, \\
&\text{Ogden \cite{Ogden1972_142d}: }\nonumber\\
&\quad \en = \brac{\lambda_1 ^{1.3} + \lambda_2 ^{1.3} + \lambda_3 ^{1.3} -3}+ 1.5\brac{J-1} ^{2}\,.
\end{align}
\end{subequations}
Here, $\lambda_c = \sqrt{\iIi/3}$, $\beta_c=\mathcal{L} ^{-1}\fOf{\lambda_c/\sqrt{N_c}}$, and $\mathcal{L}^{-1}$ denotes the inverse Langevin function.
The constants are set to $N_c=28$, corresponding to the number of polymer chain segments, and $c_{AB}\approx3.7910$, where the latter offsets the energy density to zero at $\F=\I$, since the Arruda--Boyce formulation does not vanish in the undeformed configuration.

In the anisotropic cases, the considered ground-truth behaviors are as follows:
\begin{subequations}
\begin{align}
     & \text{Anisotropic Neohookean (AN) \cite{Ning2006_6dc7}: } \nonumber \\
     & \quad \en = \brac{\iIi-3} + \mac{\iIVi-1}^{2} + 1.5\brac{J-1} ^{2}\,, \\
     & \text{Holzapfel--Gasser--Ogden (HGO)*\cite{Holzapfel2000_7a2d}: } \nonumber \\
     & \quad \en = \brac{\iIi-3} + 0.25\brac{\exp{2\mac{\iIVi-1} ^{2}}-1} + 1.5\brac{J-1} ^{2}\,, \\
     & \text{Meaney \cite{Meaney2003_534c}: } \nonumber \\
     & \quad \en = 0.5 \brac{\iIi-3} + 0.5 \brac{\iIVi ^{2} + 2 \iIVi ^{-1} -3} + 1.5\brac{J-1} ^{2}\,, \\
     & \text{Merodio--Ogden (MO)\cite{Merodio2005_4c0c}: } \nonumber \\
     & \quad \en = 0.5 \brac{\iIi-3} + 0.5 \brac{\iVi-1}^{2} + 1.5\brac{J-1} ^{2}\,, \\
     & \text{Humphrey--Yin (HY)*\cite{Humphrey1987_2c2a}: } \nonumber \\
     & \quad \en = 0.1 \brac{\exp{5 \brac{\iIi -3}}-1} + \brac{\exp{8 \mac{\sqrt{\iIVi}-1}^{2}}-1}+ 2.5\brac{J-1} ^{2}\,, \\
     & \text{Gasser--Ogden--Holzapfel (GOH)*\cite{Gasser2006_3595}: } \nonumber \\
     & \quad \en = 0.5 \brac{\iIi-3}
    +  \frac{1}{6}\brac{\exp{3 \mac{\frac{\iIi}{6}+\frac{\iIVi}{2}-1} ^{2}}-1}
    + 2.5\brac{J-1} ^{2}\,.
\end{align}
\end{subequations}
Moreover, without loss of generality, the structural vector is taken to be homogeneously aligned in the positive $y$-direction, i.e. $ \StrucVec = \brac{0 \quad 1 \quad 0}^{T}.$ 
This benchmark set is chosen deliberately to include both ground-truth laws that are inside the function space represented by the chosen CANN architecture and laws containing invariants or functional forms outside that function space. 
This enables separate assessment of exact term recovery and out-of-function-space approximation.
For further details on synthetic data generation, we refer the interested reader to \tab~\ref{tab:parameters} in \ref{sec:hyperparameters}.

\subsubsection{Error metric definitions}

During comparison of ground-truth and discovered behavior, we quantify model performance using the \textit{coefficient of determination}, commonly referred to as the $R^2$-score.
This metric is defined as
\begin{equation}
    R^2 := 1 - \frac{\sum_{i=1}^{n}\left(y_i^{\text{true}} - y_i^{\text{pred}}\right)^2}{\sum_{i=1}^{n}\left(y_i^{\text{true}} - \bar{y}\right)^2}\,, 
    \qquad 
    \bar{y} := \frac{1}{n}\sum_{i=1}^{n} y_i^{\text{true}}\,.
\end{equation}
Here, $n$ denotes the number of data points, and $y$ represents the quantity of interest, such as the strain-energy density or a component of the stress tensor.
The values $y_i^{\text{true}}$ and $y_i^{\text{pred}}$ correspond to the true and predicted responses, respectively, while $\bar{y}$ denotes the mean of the true values.

The coefficient of determination is a global metric in the sense that it summarizes model performance over an entire dataset with a single scalar value.
However, it is also useful to assess performance locally, i.e. to quantify the prediction error for an individual input.
For this purpose, we use the \textit{normalized error}, defined as
\begin{equation}
    \epsilon_{\text{norm}} := \frac{\abs{y^{\text{true}}-y^{\text{pred}}}}{\text{median}(\abs{y^{\text{true}}})}\,,
    \label{eq:err-def}
\end{equation}
where $\text{median}\fOf{\bullet}$ denotes the median\footnote{The median is used instead of the mean to define a robust normalization scale, since several benchmark models contain exponential terms that generate large outlying values for selected inputs.}
of some set of values $\bullet$ in the training dataset.
When the value of interest is a tensor, e.g. the Kirchhoff stress, then we take the median of the norm of that tensor in the training dataset.

\section{Results}
\label{sec:benchmark}

The results are organized to progress from interpretable, path-wise comparisons to stringent tests of generalization and practical predictive performance.
We first consider sparse term recovery and path-wise agreement in \sect~\ref{sec:sparse-term-recovery}, where the discovered constitutive models are compared with the ground truth along canonical loading paths; this reveals whether the CANN identifies the correct active terms and reproduces the associated stress response in mechanically interpretable settings.
We then examine generalization in multi-dimensional deformation space in \sect~\ref{sec:multidimensional}, motivated by the fact that one-dimensional loading paths probe only a small subset of the admissible deformation states; this analysis therefore assesses accuracy over broader domains and distinguishes performance on deformation states that are seen versus unseen during training.
Finally, we perform an independent forward finite element validation in \sect~\ref{sec:forward-simulation} on an unseen geometry and loading configuration to evaluate how the discovered models perform in a practically relevant setting.
Collectively, these subsections assess interpretability, generalization, and predictive utility.

\subsection{Sparse term recovery and path-wise agreement}
\label{sec:sparse-term-recovery}

\begin{figure}[!tb]
	\begin{center}
		\includegraphics[width=.95\textwidth]{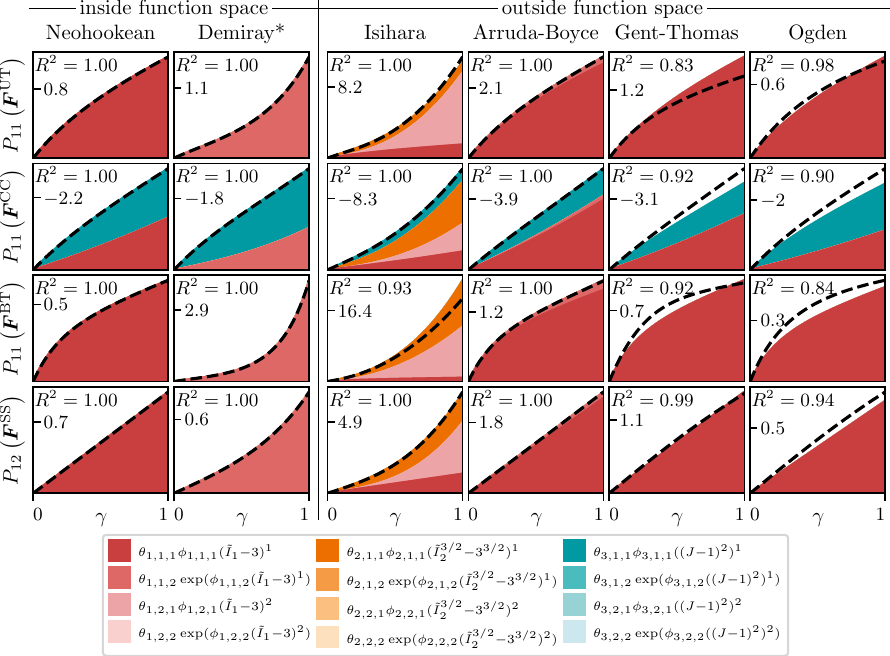}
	\end{center}
	\caption{\textbf{Comparison of discovered isotropic models first Piola-Kirchhoff stress against ground truth.}
    Relevant components of the first Piola-Kirchhoff stress for uniaxial tension (UT), confined compression (CC), biaxial tension (BT), and simple shear (SS),  as defined in \eqs~\ref{eq:pths-1} and \ref{eq:pths-2}, are shown in rows 1--4, respectively, while the Neohookean, Demiray, Isihara, Arruda-Boyce, Gent-Thomas, and Ogden models are shown in columns 1--6, respectively.
    The ground-truth response is shown as a dashed black line, whereas the discovered response is presented as a stacked area plot, with colors indicating the contribution of individual terms.
    The coefficient of determination ($R^{2}$-score) is shown in the top left of each plot. 
    Models annotated with ``*'' contain exponential terms in the ground truth.
}
	\label{fig:comp-stress-0}
\end{figure}

\begin{figure}[!tb]
	\begin{center}
		\includegraphics[width=.95\textwidth]{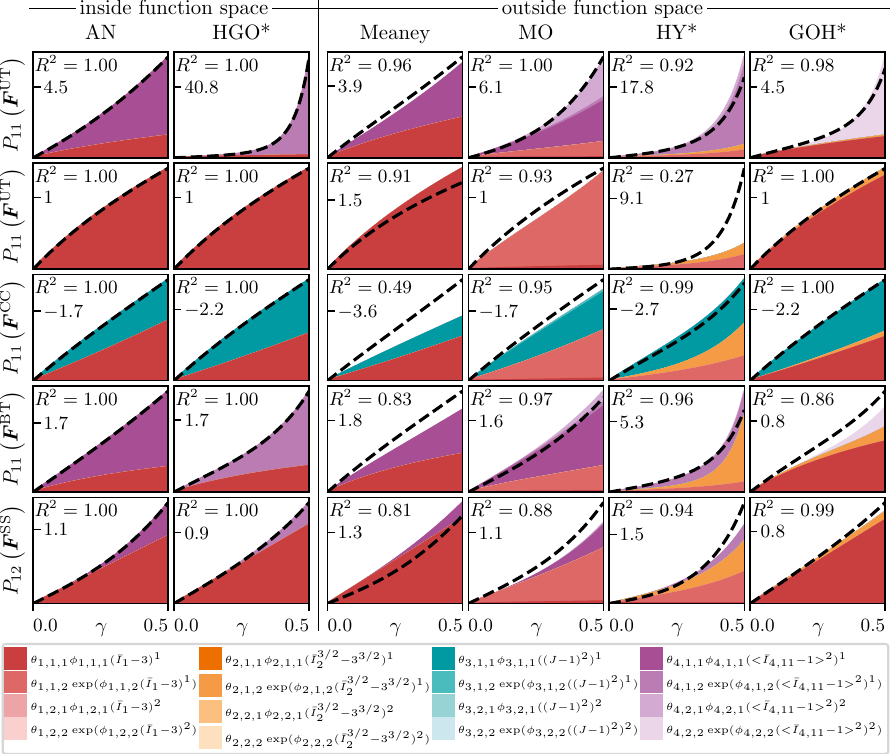}
	\end{center}
	\caption{\textbf{Comparison of discovered anisotropic models stress against ground truth.}
        The first Piola--Kirchhoff stress for uniaxial tension (UT) with $\StrucVec=\brac{1\quad 0 \quad 0}^{T}$, UT with $\StrucVec=\brac{0\quad 1 \quad 0}^{T}$, confined compression (CC) with $\StrucVec=\brac{1\quad 0 \quad 0}^{T}$, biaxial tension (BT) with $\StrucVec=\brac{1\quad 0 \quad 0}^{T}$, and simple shear (SS) with $\StrucVec=\brac{0\quad 1 \quad 0}^{T}$, are shown in rows 1--5, respectively, while the Anisotropic Neohookean (AN), Holzapfel--Gasser--Ogden (HGO), Meaney, Merodio--Ogden (MO), Humphrey--Yin (HY), and Gasser--Ogden--Holzapfel (GOH) models are shown in columns 1--6, respectively.
    The ground-truth behavior is shown as a single dashed black line, whereas the discovered behavior is presented as a stacked area plot with colors indicating the contribution of individual terms as specified in the legend.
    Additionally, the coefficient of determination ($R^{2}$-score) between the true and discovered behavior for each model and loading path is shown in the top-left corner of each corresponding plot.
    Models annotated with ``*'' contain exponential terms in the ground truth.
    }
	\label{fig:comp-trans-stress-0}
\end{figure}

The stress responses of the discovered models are compared to the isotropic and anisotropic ground-truth models in \figs~\ref{fig:comp-stress-0} and \ref{fig:comp-trans-stress-0}, respectively, for various loading paths. 
The considered loading paths include uniaxial tension (UT), confined compression (CC), biaxial tension (BT), and simple shear (SS), which are defined, respectively, as
\begin{subequations}
\begin{align}
\F^{\text{UT}} \fOf{\gamma} & = \begin{bmatrix}
1+\gamma & 0 & 0 \\
0 & \frac{1}{\sqrt{1+\gamma }} & 0 \\
0 & 0 & \frac{1}{\sqrt{1+\gamma }} 
\end{bmatrix}\,, &
\F^{\text{CC}} \fOf{\gamma} & = \begin{bmatrix}
\frac{1}{1+\gamma} & 0 & 0 \\
0 & 1 & 0 \\
0 & 0 & 1
\end{bmatrix}\,, \label{eq:pths-1} \\
\F^{\text{BT}} \fOf{\gamma} & = \begin{bmatrix}
1+\gamma & 0 & 0 \\
0 & 1+\gamma & 0 \\
0 & 0 & \frac{1}{\brac{1+\gamma }^{2}} 
\end{bmatrix}\,, &
\F^{\text{SS}} \fOf{\gamma} & = \begin{bmatrix}
1 & \gamma & 0 \\
0 & 1 & 0 \\
0 & 0 & 1
\end{bmatrix}\,.
\label{eq:pths-2}
\end{align}
\label{eq:pths}
\end{subequations}

These loading paths are shown in rows 1--4 of the isotropic comparison in \fig~\ref{fig:comp-stress-0}, corresponding to UT, CC, BT, and SS, respectively.
The columns show the Neohookean, Demiray, Isihara, Arruda--Boyce, Gent--Thomas, and Ogden models.
For the anisotropic benchmarks in \fig~\ref{fig:comp-trans-stress-0}, we use the same deformation gradients defined in \eqs~\eqref{eq:pths}, but vary the structural vector relative to the loading axes.
Rows 1 and 2 show UT with $\StrucVec=\brac{1\quad 0 \quad 0}^{T}$ and $\StrucVec=\brac{0\quad 1 \quad 0}^{T}$, respectively, corresponding to loading parallel and transverse to the structural direction.
Rows 3 and 4 show CC and BT with $\StrucVec=\brac{1\quad 0 \quad 0}^{T}$, while row 5 shows SS with $\StrucVec=\brac{0\quad 1 \quad 0}^{T}$.

\clearpage

Overall, the discovered models exhibit excellent agreement with the ground truth (shown with a dashed line), with coefficients of determination $R^{2}$ exceeding 0.95 in the majority of cases (\figs~\ref{fig:comp-stress-0} and \ref{fig:comp-trans-stress-0}).

\textbf{Ground-truth laws that are in the CANN function space are near-exactly recovered.}
In the isotropic setting this is observed for the Neohookean and Demiray models where the isochoric terms \smallmath{\theta_{1,1,1}\phi_{1,1,1}\brac{\iIi-3}} and \smallmath{\theta_{1,1,2}\,\exp{\phi_{1,1,2}\brac{\iIi-3}}} are correctly identified, respectively, and the volumetric term \smallmath{\theta_{3,1,1}\phi_{3,1,1}\brac{J-1}^{2}} is correctly identified for both models (\fig~\ref{fig:comp-stress-0} columns 1 and 2, respectively).
In the anisotropic case, this was demonstrated for the 
anisotropic Neohookean (AN) and Holzapfel-Gasser-Ogden (HGO) ground-truth models where the CANN correctly identifies the anisotropic terms \smallmath{\theta_{4,1,1}\phi_{4,1,1}\mac{\iIVi-1}^{2}} and \smallmath{\theta_{4,1,2}\,\exp{\phi_{4,1,2}\mac{\iIVi-1}^{2}}}, respectively, as well as the isotropic terms \smallmath{\theta_{1,1,1}\phi_{1,1,1}\brac{\iIi-3}} and  \smallmath{\theta_{3,1,1}\phi_{3,1,1}\brac{J-1}^{2}} for both models (\fig~\ref{fig:comp-trans-stress-0} columns 1 and 2, respectively).
The recovery of exponential terms is particularly important, because the corresponding material parameter appears inside a nonlinear function; this goes beyond the standard EUCLID formulation, which identifies linear coefficients of fixed library terms but does not directly calibrate parameters embedded inside nonlinear functions \cite{Flaschel2021_4e47}.

\textbf{Shared terms are recovered even when the full ground-truth law is not representable by the chosen CANN function space.}
For example, in the isotropic setting, the Isihara model and chosen CANN architecture both contain the terms \smallmath{\brac{\iIi - 3}}, \smallmath{\brac{\iIi - 3}^{2}}, and \smallmath{\brac{J - 1}^{2}}, but the Isihara model also contains the term \smallmath{\brac{\iIIi-3}} which is not contained in the CANN function space.
Nevertheless, the terms \smallmath{\brac{\iIi - 3}}, \smallmath{\brac{\iIi - 3}^{2}}, and \smallmath{\brac{J - 1}^{2}} are correctly identified by the CANN.
A similar pattern is observed in the anisotropic setting: in the Meaney, Merodio--Ogden (MO), and Gasser--Ogden--Holzapfel (GOH) models, CANN--EUCLID correctly identifies the term \smallmath{\theta_{1,1,1}\phi_{1,1,1}\brac{\iIi-3}}, while in the Humphrey--Yin (HY) model it correctly identifies the term \smallmath{\theta_{1,1,2}\exp{\phi_{1,1,2}\brac{\iIi-3}}}.
This occurs despite the fact that these models contain additional terms that are not present in the chosen CANN function space.
The only exception is the MO model: although \smallmath{\brac{\iIi -3}} is the only \smallmath{\iIi}-based term in the ground-truth law and is correctly identified, CANN--EUCLID also activates the additional term \smallmath{\theta_{1,1,2}\exp{\phi_{1,1,2}\brac{\iIi-3}}} in the discovered MO behavior.
Apart from this exception, all \smallmath{\iIi}- and \smallmath{J}-based terms that are present in the CANN function space but absent from the Meaney, MO, GOH, and HY ground-truth models are correctly omitted where appropriate.

\textbf{Ground-truth terms that are absent from the CANN function space are typically approximated by other terms.}
This is exemplified by the Isihara model, which contains the non-polyconvex term \smallmath{\brac{\iIIi-3}}.
In this case, CANN--EUCLID instead identifies the polyconvex term \smallmath{\theta_{2,1,1}\phi_{2,1,1}\brac{\iIIi^{3/2}-3^{3/2}}}, resulting in comparable path-wise behavior (\fig~\ref{fig:comp-stress-0}, column 3).
Moreover, complex functional forms, such as logarithmic terms, hyperbolic sine functions, and principal stretches raised to non-integer powers, are approximated by simpler terms contained in the chosen CANN function space.
This is demonstrated by CANN--EUCLID's ability to approximate the response of the Gent--Thomas, Arruda--Boyce, and Ogden models with just two terms: \smallmath{\theta_{1,1,1}\phi_{1,1,1}\brac{\iIi-3}} and \smallmath{\theta_{3,1,1}\phi_{3,1,1}\brac{J-1}^{2}} (\fig~\ref{fig:comp-stress-0}, columns 4--6).
Similarly, all anisotropic $\iIVi$- and $\iVi$-based terms in the Meaney, MO, HY, and GOH models are absent from the chosen CANN function space.
Nevertheless, CANN--EUCLID approximates the anisotropic stress contributions with good accuracy in loading cases where these terms are active, i.e., UT with $\StrucVec=\brac{1 \quad 0 \quad 0}^{T}$, BT, and SS.
For the Meaney model, the term \smallmath{\brac{\iIVi^{2}+2\iIVi^{-1}-3}} is approximated by \smallmath{\theta_{4,1,1}\phi_{4,1,1}\mac{\iIVi-1}^{2}} (\fig~\ref{fig:comp-trans-stress-0}, column 3).
For the MO model, the term \smallmath{\brac{\iVi-1}^{2}} is approximated by a combination of \smallmath{\theta_{4,1,1}\phi_{4,1,1}\mac{\iIVi-1}^{2}}, \smallmath{\theta_{4,1,2}\exp{\phi_{4,1,2}\mac{\iIVi-1}^{2}}}, and \smallmath{\theta_{4,2,1}\phi_{4,2,1}\mac{\iIVi-1}^{4}} (\fig~\ref{fig:comp-trans-stress-0}, column 4).
For the HY model, the term \smallmath{\brac{\exp{8 \mac{\sqrt{\iIVi}-1}^{2}}-1}} is approximated by \smallmath{\theta_{4,1,2} \exp{\phi_{4,1,2}\mac{\iIVi-1}^{2}}} and \smallmath{\theta_{4,2,1}\phi_{4,2,1}\mac{\iIVi-1}^{4}} (\fig~\ref{fig:comp-trans-stress-0}, column 5).
Finally, for the GOH model, the term \smallmath{\frac{1}{6}\brac{\exp{3 \mac{\frac{\iIi}{6}+\frac{\iIVi}{2}-1}^{2}}-1}} is approximated by \smallmath{\theta_{4,2,1}\phi_{4,2,1}\mac{\iIVi-1}^{4}} (\fig~\ref{fig:comp-trans-stress-0}, column 6).

\subsection{Generalization in multi-dimensional deformation space}
\label{sec:multidimensional}

\begin{figure}[!tb]
	\begin{center}
		\includegraphics[width=.99\textwidth]{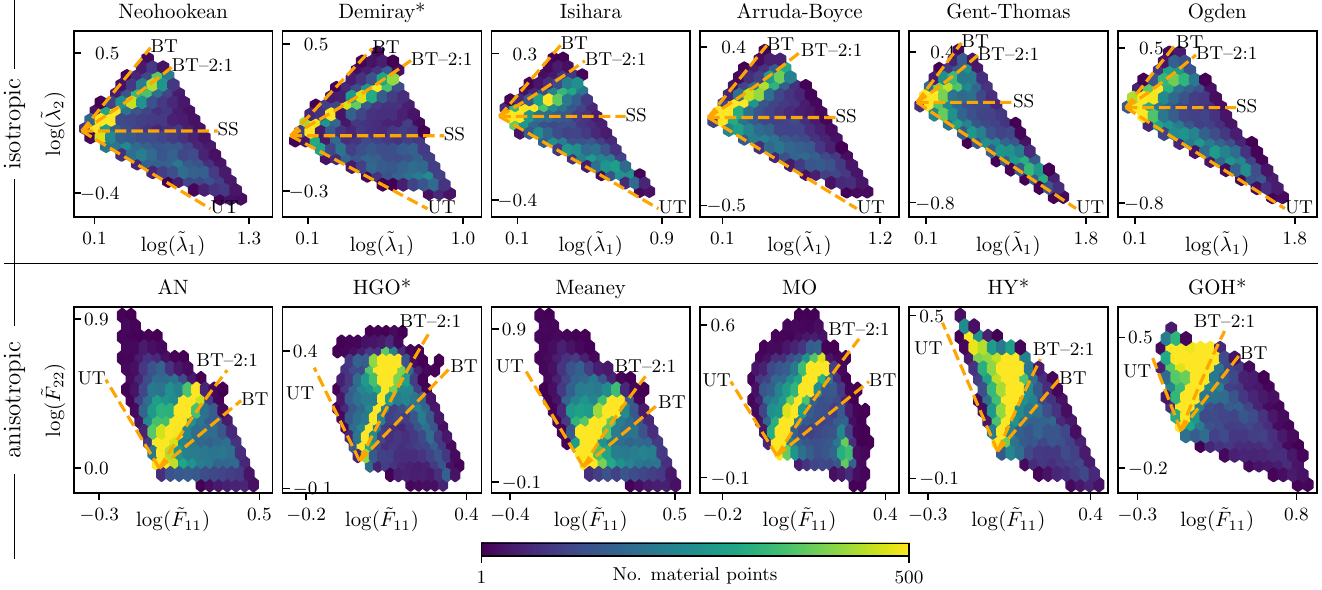}
	\end{center}
	\caption{\textbf{Deformation states seen in training data}.
        In the isotropic cases (top), the strain-energy density function and principal stresses can be written in terms of the principal stretches.
        The frequency of the first and second isochoric principal stretches, $\iso{\lambda}_1$ and $\iso{\lambda}_2$, in the training data is shown for the Neohookean, Demiray, Isihara, Arruda--Boyce, Gent--Thomas, and Ogden models in columns 1--6, respectively.
        In the anisotropic cases (bottom), the strain-energy density function and principal stresses cannot, in general, be written solely in terms of the principal stretches.
        We therefore plot the frequency of the $\iso{F}_{11}$ and $\iso{F}_{22}$ components of the isochoric deformation gradient.
        The anisotropic Neohookean (AN), Holzapfel--Gasser--Ogden (HGO), Meaney, Merodio--Ogden (MO), Humphrey--Yin (HY), and Gasser--Ogden--Holzapfel (GOH) models are shown in columns 1--6, respectively.
        The paths corresponding to uniaxial tension (UT), equibiaxial tension (BT), and simple shear (SS), as defined in \eqs~\eqref{eq:pths}, are indicated by dashed orange lines; the simple-shear path is not shown in the anisotropic case because it degenerates to a point in this projection.
        The line corresponding to biaxial tension with a 2:1 ratio, i.e., the loading applied to the training sample in \fig~\ref{fig:geoms}, is indicated by the annotation ``BT--2:1''.
        The figure illustrates that a single heterogeneous test in the unsupervised paradigm induces deformation states associated with multiple conventional supervised loading paths.
        It also shows that the resulting dataset is imbalanced, with most deformation states clustering around the BT--2:1 line.
     }
	\label{fig:training-deformations}
\end{figure}

Observing the discovered behavior along distinct loading paths as shown in \figs~\ref{fig:comp-stress-0} and \ref{fig:comp-trans-stress-0} is useful for elucidating which terms contribute to the behavior and for providing a cursory comparison between the ground-truth and discovered behavior.
However, evaluating the performance of the discovered behavior in this manner has at least two notable limitations:

\textit{(i) One-dimensional lines through a six-dimensional deformation space.}
\begin{adjustwidth}{12pt}{}
    General three-dimensional hyperelastic behavior depends on six independent variables, i.e., the six independent components of the left or right Cauchy--Green tensor.
    The loading paths described in \eqs~\eqref{eq:pths} and used for \figs~\ref{fig:comp-stress-0} and \ref{fig:comp-trans-stress-0} define one-dimensional lines through this deformation space.
    These lines necessarily omit large regions of the six-dimensional domain that describes the possible deformation state at a material point.
    As such, evaluating the behavior on these one-dimensional paths alone is not sufficient, as it excludes large regions of deformation space that the discovered models may encounter during deployment, e.g., in a finite element solver.
    This is illustrated in \fig~\ref{fig:training-deformations}, where uniaxial tension (UT), biaxial tension (BT), and simple shear (SS) are shown as dashed orange lines overlaid on frequency plots of the deformation states observed in the training data for each ground-truth model.
    Since the full six-dimensional space cannot be visualized directly, these paths are projected onto two-dimensional manifolds corresponding to the isochoric principal stretches in the isotropic cases and to two components of the deformation gradient in the anisotropic cases.
    These projected paths occupy only small regions of the sampled deformation space, highlighting domains in which the accuracy of the discovered models is not assessed by path-wise comparisons alone.
\end{adjustwidth}

\begin{figure}[tp]
	\begin{center}
		\includegraphics[width=.99\textwidth]{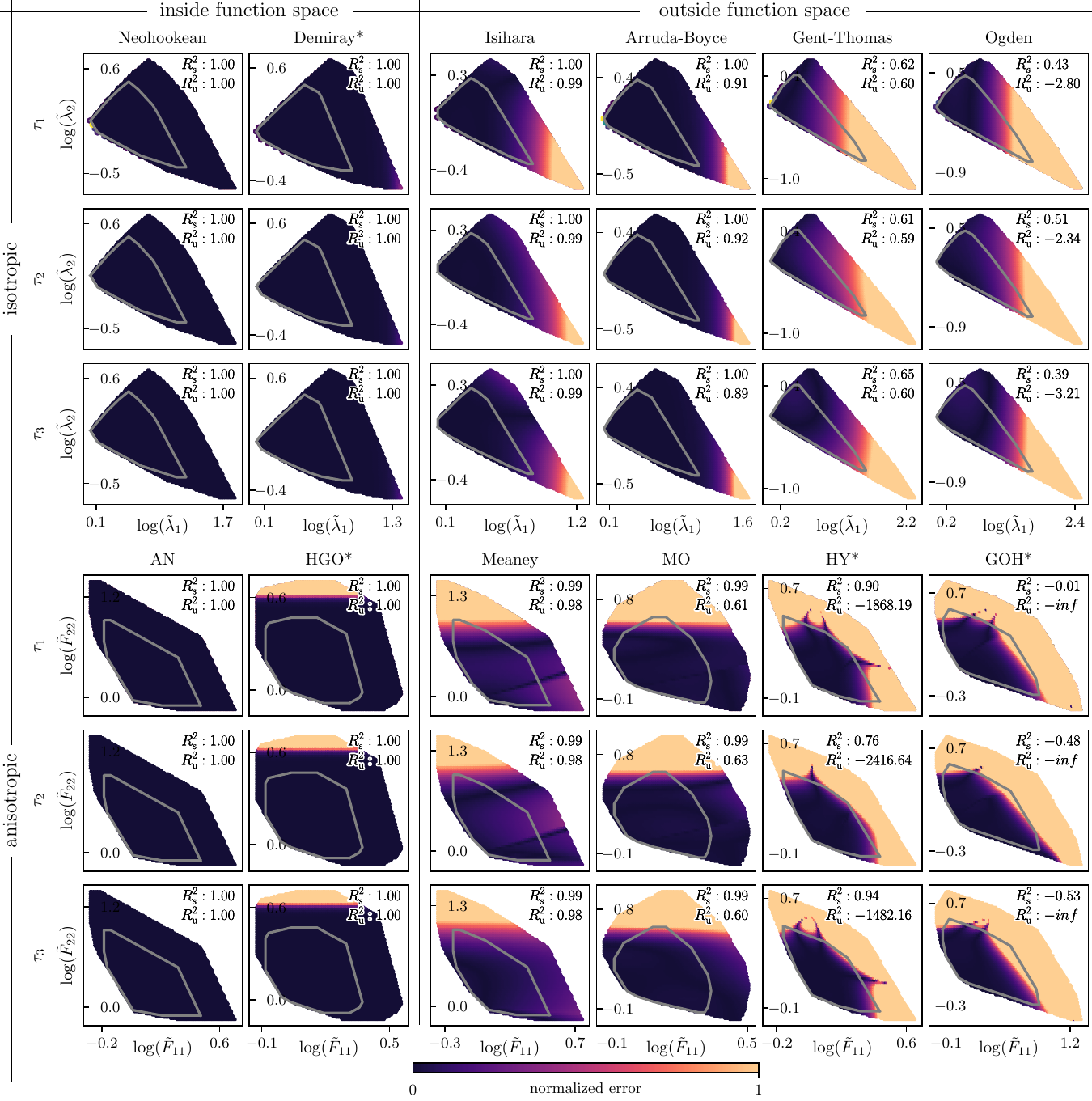}
	\end{center}
	\caption{\textbf{Accuracy of discovered CANNs evaluated in isochoric deformation space.}
        The normalized error of the first, second, and third principal Kirchhoff stresses discovered by the CANN as compared to the ground truth are displayed in rows 1--3, respectively, for both the isotropic (top) and anisotropic (bottom) cases, with results for different material models grouped in columns.
        Plots in the isotropic cases (top) use the log of the first and second isochoric principal stretches for the $x$-- and $y$--axes respectively;
        the third isochoric principal stretch is implicitly defined through $\tilde{\lambda}_3 = 1/\brac{\tilde{\lambda}_1\tilde{\lambda}_2}$.
        In the anisotropic cases (bottom), the $\iso{F}_{11}$ and $\iso{F}_{22}$ components of the isochoric deformation gradient are used for the axes. 
        Moreover, the sampled deformation gradients are chosen to be diagonal with $\iso{F}_{33} = 1/\brac{\iso{F}_{11}\iso{F}_{22}}$ and the direction of the structural vector is taken to be $\StrucVec = \brac{0\quad 1 \quad 0}^{T}$.
        The convex hull of the training data (visible in \fig~\ref{fig:training-deformations}) is indicated by a grey line and the evaluation domain has been extended beyond the convex hull to assess the generalizability of the discovered CANNs.
        The $R^2$-scores for the seen data $R^{2}_\text{s}$ and unseen data $R^{2}_\text{u}$ are reported in the top-right corner of each plot. 
    }
	\label{fig:princ-stretch-space}
\end{figure}

\clearpage
\textit{(ii) Lack of discrimination between predictions on seen and unseen deformation states.}
\begin{adjustwidth}{12pt}{}
    In supervised model calibration or discovery using, for example, uniaxial test data, the seen deformation states can be defined naturally from the maximum stretch reached during training.
    Generalization can then be assessed by evaluating the discovered model at stretches beyond those observed during training.
    In the context of the loading paths used in \figs~\ref{fig:comp-stress-0} and \ref{fig:comp-trans-stress-0}, this would amount to choosing a value of $\gamma$ corresponding to the maximum stretch seen during training.
    However, in the unsupervised case, the deformation states in the training data are diverse and heterogeneous.
    It is therefore not immediately clear which range of $\gamma$ values should be considered seen during training and which should be considered unseen.
\end{adjustwidth}
To address these points, we assess the accuracy of the discovered behavior on higher-dimensional domains of the possible deformation space that are seen and unseen during training in \fig~\ref{fig:princ-stretch-space}.
Since effectively visualizing a three- or six-dimensional domain is intractable on a two-dimensional page or screen, we choose to assess the behavior on two-dimensional manifolds through the six-dimensional deformation space.
The normalized error, as defined in \eq~\eqref{eq:err-def}, in the first, second, and third principal Kirchhoff stresses, are shown in rows 1--3, respectively, for the isotropic cases (top) and anisotropic cases (bottom) of \fig~\ref{fig:princ-stretch-space}.
In the case of isotropic behavior, we use the manifold defined by isochoric deformations as done in \fig~\ref{fig:training-deformations}; i.e. we use \smallmath{\iso{\lambda}_1} and \smallmath{\iso{\lambda}_2} and set \smallmath{\iso{\lambda}_3 = 1/\brac{\iso{\lambda}_1\iso{\lambda}_2}}.
In the anisotropic case, we restrict ourselves to visualizing the $\iso{F}_{11}$ and $\iso{F}_{22}$ components of the isochoric deformation gradient whilst noting that this is only a slice through the six-dimensional domain of independent variables. 
Moreover, the sampled deformation gradients are chosen to be diagonal with \smallmath{\iso{F}_{33} = 1/\brac{\iso{F}_{11}\iso{F}_{22}}} and the direction of the structural vector is taken to be $\StrucVec = \brac{0\quad 1 \quad 0}^{T}$.

\begin{figure}[tb]
	\begin{center}
		\includegraphics[width=\textwidth]{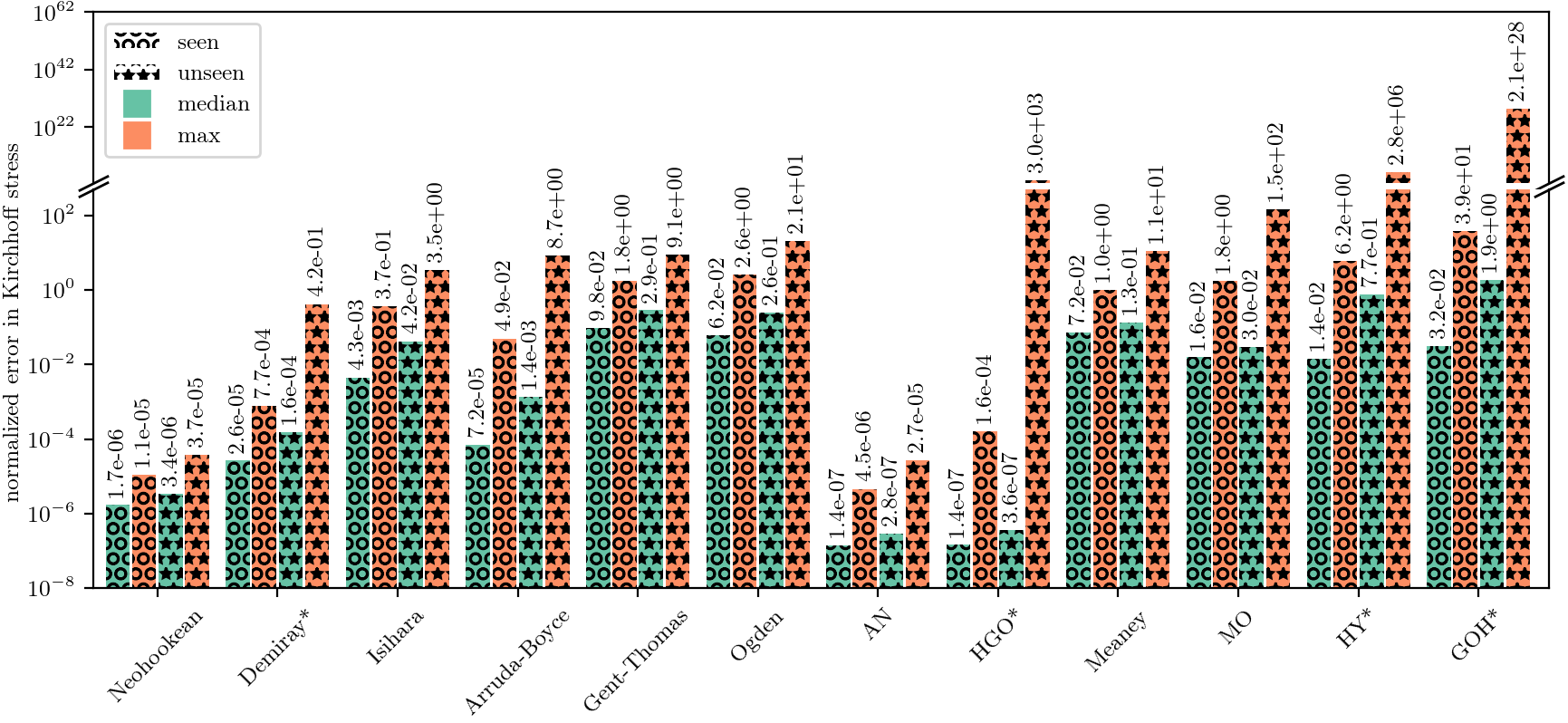}
	\end{center}
	\caption{\textbf{Statistics for CANNs stress-prediction accuracy on seen and unseen deformations.} 
        The median and maximum of the normalized errors in the Kirchhoff principal stresses presented in \fig~\ref{fig:princ-stretch-space} are shown for the seen and unseen domains of deformation for the isotropic and anisotropic cases.
}
	\label{fig:iso-err-stats}
\end{figure}

The sampled domains are chosen to evaluate the performance and generalizability of the models on seen and unseen deformation states.
The domains are chosen by obtaining the convex hull of the deformation states used during training as shown in \fig~\ref{fig:training-deformations}.
The sampling domain is then obtained by simply scaling the points on the perimeter of the convex hull by a factor $c_{\text{gen}}$ ($1<c_{\text{gen}}$).
We make use of log values since this places the undeformed location at the origin, which allows for increasing the sampling domain by simply scaling the points without translating the undeformed location\footnote{Scaling compressive stretches in the log space results in more compressive stretches ($\log(\tilde{\lambda}_2)<0 \Rightarrow c_{\text{gen}}\log(\tilde{\lambda}_2)<\log(\tilde{\lambda}_2)$), and similarly for tensile stresses ($0< \log(\tilde{\lambda}_1)\Rightarrow \log(\tilde{\lambda}_1)<c_{\text{gen}}\log(\tilde{\lambda}_1)$).
If the log of the stretches was not used, scaling would result in all stretches becoming more tensile ($\tilde{\lambda}_1 < c_{\text{gen}}\tilde{\lambda}_1 $).}.
The larger the value of $c_{\text{gen}}$ the larger the domain of unseen deformation states; in this work, we choose $c_{\text{gen}}=1.5$.
The convex hull of the seen deformation states is shown as a grey line in \fig~\ref{fig:princ-stretch-space} and the associated sampled domains are clear from the colored areas.
Additionally, the coefficients of determination on the seen $R_s^{2}$ and unseen $R_u^{2}$ data are shown in the top-right corner of each normalized error plot.
To aid in the analysis and discussion of \fig~\ref{fig:princ-stretch-space}, the median and maximum of the stress prediction errors in the seen and unseen deformation domains are presented as a bar plot for each model in \fig~\ref{fig:iso-err-stats}.

\bigskip
\textbf{Exceptional performance on seen and unseen deformation states is obtained when the ground-truth models are recovered near-exactly.}
This is demonstrated by the Neohookean and Demiray models in the isotropic case and the anisotropic Neohookean (AN) and Holzapfel--Gasser--Ogden (HGO) models in the anisotropic cases, all of which were shown to be captured exactly in \sect~\ref{sec:sparse-term-recovery}, \figs~\ref{fig:comp-stress-0} and \ref{fig:comp-trans-stress-0}.
Here, these models all achieve $R^{2}$-scores of $1.00$ for all three principal stresses on seen and unseen deformations (\fig~\ref{fig:princ-stretch-space}, columns 1 and 2, Anisotropic and Isotropic).
Additionally, the highest normalized error between all of them on seen deformation states is below $7.7 \times 10 ^{-4}$ (\fig~\ref{fig:iso-err-stats}).
However, despite the near-exact recovery of the ground-truth models, and median errors on unseen deformations being below $1.6 \times 10 ^{-4}$ (\fig~\ref{fig:iso-err-stats}),
the Demiray and HGO models, i.e. the models containing exponentials, still result in maximum normalized errors of $0.42$ and $4.4 \times 10 ^{3}$ on the unseen deformation domain, respectively.
This highlights the sensitivity of exponential-based models and provides a first indication of the challenges surrounding generalizability in the presence of such terms.

\textbf{The median error in the seen deformation domain is low across all models}.
This holds true even in cases where the ground truth contains a pseudo-invariant not used in the CANN function space, e.g. the Isihara and Merodio--Ogden (MO) models, which have median errors of $4.3 \times 10 ^{-3}$ and $1.8 \times 10 ^{-2}$, respectively (\fig~\ref{fig:iso-err-stats}).
It also holds when the ground truth has a substantially different functional form than the activation functions in the CANN, e.g. the Ogden and the Meaney models, with median errors of $6.2 \times 10 ^{-2}$ and $4.9 \times 10 ^{-2}$, respectively (\fig~\ref{fig:iso-err-stats}).
Moreover, regions of the seen domain with more training data points tend to exhibit lower errors than sparsely sampled regions, such as those close to the boundary of the seen domain.
For example, the errors for the HY and GOH models in \fig~\ref{fig:princ-stretch-space} are much larger close to the convex hull of the seen deformation domain; these regions correspond to deformation states reached by fewer material points in \fig~\ref{fig:training-deformations}.
The low number of data points in these deformation states gives them a smaller relative contribution to the loss function, which likely contributes to the larger normalized errors observed in these regions.
This suggests a practical avenue for increasing the accuracy of the discovered model in deformation regimes of interest: design heterogeneous tests that drive more material points into those regimes.

\textbf{Generalization performance is highly dependent on the presence or absence of exponential terms}.
This is clearest for the HY and GOH models, both of which contain exponential terms that are not contained in the chosen CANN function space.
In these cases, the median errors in the unseen domain are $0.41$ and $2.2$, respectively.
By contrast, the median errors in the unseen domain for the Meaney and MO cases are $0.11$ and $0.043$, respectively.
These results indicate that reliable extrapolation cannot be assumed when the sampled deformation states do not sufficiently probe a potential strain-stiffening regime that governs the material response outside the training domain.

\subsection{Independent forward finite element validation}
\label{sec:forward-simulation}

\begin{figure}[!tb]
	\begin{center}
		\includegraphics[width=.99\textwidth]{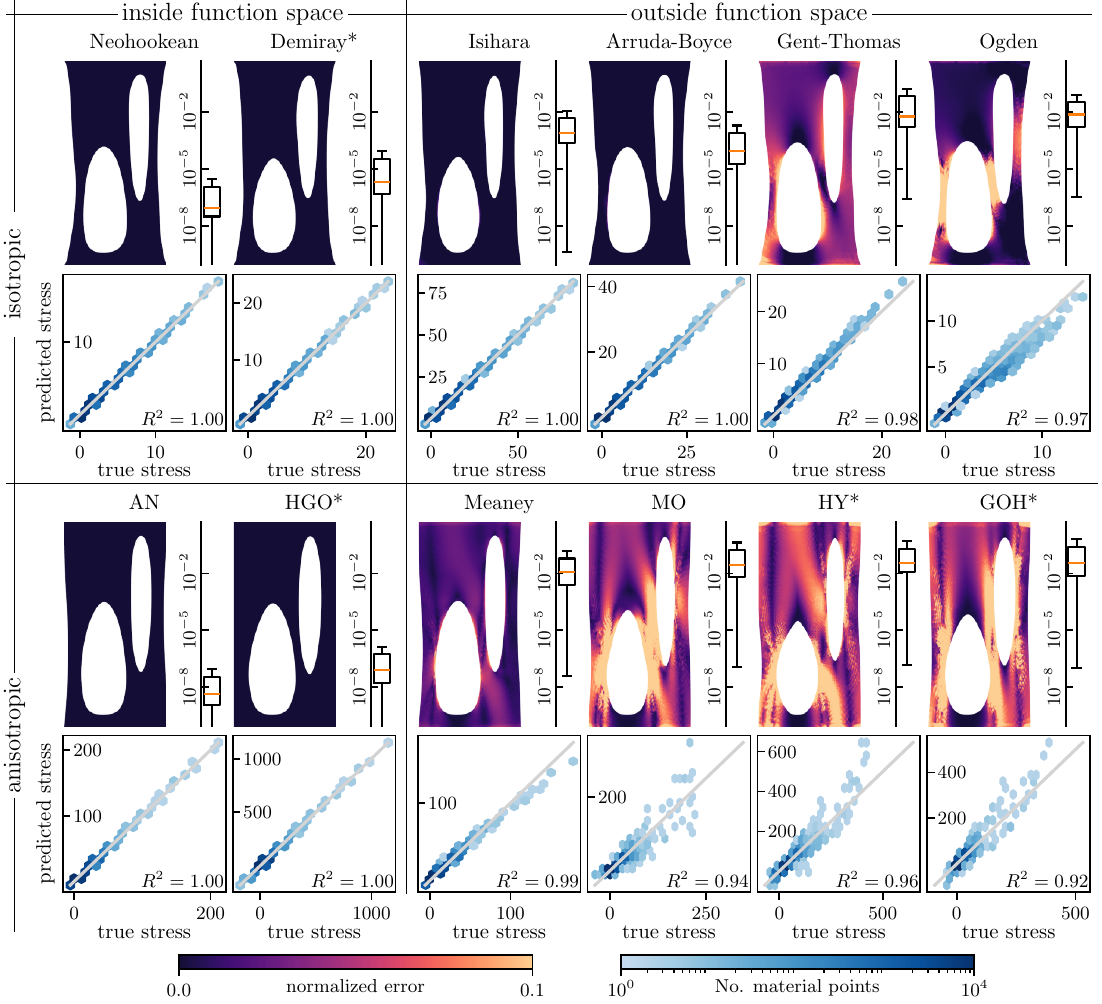}
	\end{center}
	\caption{\textbf{Comparison of ground-truth and discovered stress fields in validation simulations.}
        Results for the isotropic models are grouped above the results for the anisotropic models, and the results for the models inside the function space of the CANN are grouped on the left.
        The normalized error in the Kirchhoff stress defined in \eq~\eqref{eq:err-def} is shown on the deformed configuration in the top row of each group next to boxplots indicating the distribution of the normalized error 
        and parity plots of principal Kirchhoff stresses are shown in the bottom row.
    }
	\label{fig:validation}
\end{figure}

The solution to the finite strain elasticity boundary value problem inherently minimizes the total stored elastic energy.
As such, many of the deformation states present in \fig~\ref{fig:princ-stretch-space} may be unlikely to arise in a forward FE simulation because of their excessively large associated strain-energy density values.
Hence, evaluating the accuracy of the discovered models by sampling deformation space alone, as done in \sect~\ref{sec:multidimensional}, may give a more pessimistic impression of their practical accuracy.
To estimate the accuracy that can be expected when using the discovered material models in a forward FE simulation, we perform an independent validation simulation and compare the stress field obtained with the ground-truth behavior against that obtained with the discovered behavior.
The loading and geometric parameters of the validation case are chosen to represent a challenging scenario that induces extreme deformations, as illustrated in \fig~\ref{fig:geoms}: the geometry contains elliptical holes that induce stress concentrations and is stretched by $100$~\% of its height. 
Additionally, in the anisotropic cases, the structural vector is chosen to be in the positive $y$-direction, i.e. $\StrucVec=\brac{0 \quad 1 \quad 0}^{T}$.
The results of these simulations are displayed in \figs~\ref{fig:validation}.
Isotropic models are grouped above anisotropic models, ground-truth models are grouped by column, normalized stress errors are shown spatially on the deformed geometries in the first row next to boxplots indicating the distribution of the normalized error, and the three principal Kirchhoff stresses are displayed collectively in parity plots in the second row.

\textbf{Forward validation errors are substantially lower than errors obtained from uniform sampling of the multidimensional deformation space.}
This is particularly clear for the two least accurate isotropic cases, namely the Gent--Thomas and Ogden models, for which the median normalized stress errors are only $6.0\times10^{-3}$ and $7.6\times10^{-3}$, respectively (\fig~\ref{fig:validation}, columns 5 and 6). 
These errors are more than an order of magnitude lower than the median errors on the unseen $c_{\text{gen}} \in [1.0,1.5]$ multi-dimensional deformation states from \sect~\ref{sec:multidimensional}, which were $0.29$ and $0.26$ for the Gent--Thomas and Ogden models, respectively, in \fig~\ref{fig:iso-err-stats}.
Similarly, the validation simulations for the anisotropic Meaney, HY, and GOH models yield median errors that are approximately one order of magnitude lower than those obtained from direct sampling of the deformation domain.
Median errors of $0.011$, $0.036$, and $0.037$ are observed for the validation simulations of the Meaney, HY, and GOH models, respectively 
(\fig~\ref{fig:validation}, bottom, columns 3, 5, and 6, respectively),
compared to median errors of $0.11$, $0.41$, and $2.2$ on unseen $c_{\text{gen}} \in [1.0,1.5]$ multi-dimensional deformations in \fig~\ref{fig:princ-stretch-space} (\fig~\ref{fig:iso-err-stats}).
This discrepancy arises because the statistics in \fig~\ref{fig:iso-err-stats} are computed from uniform sampling of the deformation domain and therefore do not reflect the likelihood of individual deformation states occurring in a forward simulation.
By contrast, the error statistics obtained from the validation simulations naturally account for how frequently different deformation states arise in the solution, making the resulting median errors more representative of practical FE predictions.

\textbf{Within the validation simulations, median errors remain low, while maximum errors reveal localized or dispersed model-specific discrepancies.}
In the isotropic cases, this is evident from the predicted-to-true stress parity plots, in which the predicted principal stresses lie close to the identity line for all six models (\fig~\ref{fig:validation}, top).
The coefficient of determination is $R^2=1.00$ for the Neohookean, Demiray, Isihara, and Arruda--Boyce cases, and remains high at $R^2=0.98$ and $R^2=0.97$ for the Gent--Thomas and Ogden cases, respectively (\fig~\ref{fig:validation}, top).
Moreover, the median normalized stress errors remain small across all isotropic models: below $8.3 \times 10^{-4}$ for the Neohookean, Demiray, Isihara, and Arruda--Boyce models, and below $7.6 \times 10^{-3}$ for the more challenging Gent--Thomas and Ogden models (\fig~\ref{fig:validation}, top).
Thus, even for the two least accurate isotropic validation cases, the median error remains below $1\,\%$.
The spatial error distributions further show that the larger discrepancies in the isotropic cases are localized rather than widespread.
For the Neohookean, Demiray, Isihara, and Arruda--Boyce models, the error is nearly uniformly close to zero, whereas for the Gent--Thomas and Ogden models, the larger stress errors are confined to relatively small regions near the stress concentrations around the holes (\fig~\ref{fig:validation}, top).
In the anisotropic cases, the near-exactly recovered anisotropic Neohookean (AN) and Holzapfel--Gasser--Ogden (HGO) models again display extremely low median normalized errors of $4.4 \times 10^{-9}$ and $8.2 \times 10^{-8}$, respectively (\fig~\ref{fig:validation}, bottom, columns 1 and 2, respectively).
The remaining anisotropic models also demonstrate low median errors of $0.011$, $0.028$, $0.036$, and $0.037$ for the Meaney, Merodio--Ogden (MO), Humphrey--Yin (HY), and Gasser--Ogden--Holzapfel (GOH) models, respectively (\fig~\ref{fig:validation} bottom, columns 3--6, respectively).
However, these cases also exhibit large maximum normalized errors, with values of $2.6$, $20$, $17$, and $24$ for the Meaney, MO, HY, and GOH models, respectively.
Again, larger errors tend to arise in the presence of exponential terms: the HY and GOH models contain exponential terms in the ground truth, while in the MO case the exponential term \smallmath{\theta_{4,1,2}\exp{\phi_{4,1,2}\mac{\iIVi-1}^{2}}} is included in the discovered CANN.
Moreover, these errors tend to be more spatially dispersed in the presence of exponential terms: the maximum error in the Meaney model is highly localized, whereas the larger errors in the MO, HY, and GOH cases are more spread out.

\section{Discussion}
\label{sec:discussion}
The central result of this study is that sparse, interpretable CANN-based constitutive laws can be discovered directly from full-field kinematic data and reaction forces. 
Across the isotropic and anisotropic benchmarks, CANN--EUCLID recovered the correct constitutive structure when the ground-truth law was contained in the chosen function space, produced useful approximations when it was not, and revealed clear limits of extrapolation when relevant deformation regimes were insufficiently sampled. 
These results support stress-unsupervised full-field discovery as a promising route for constitutive model identification, while also highlighting the importance of CANN architecture design, deformation-state coverage, and forward validation.

When the ground-truth constitutive law was contained in the chosen CANN function space, the framework recovered the correct active terms with near-exact accuracy. 
This was observed for the isotropic Neohookean and Demiray models and for the anisotropic Neohookean and Holzapfel--Gasser--Ogden models.
In these cases, the discovered CANNs selected the relevant isochoric, volumetric, and anisotropic terms while suppressing inactive alternatives. 
Importantly, this included exponential terms whose material parameters appear inside nonlinear functions. 
This capability goes beyond classical sparse linear-library discovery and is particularly relevant for soft biological tissues, where exponential strain-stiffening terms are common.
Moreover, for these exactly representable cases, near-exact term recovery also translated into excellent generalization over the evaluated deformation domains.

The framework also produced useful constitutive approximations when the exact ground-truth law was not representable by the selected CANN function space. 
In these cases, CANN--EUCLID generally retained shared terms that were present in both the ground truth and the CANN function space, and compensated for missing contributions using the closest available functions. 
This behavior was observed for the Isihara, Arruda--Boyce, Gent--Thomas, and Ogden models in the isotropic setting, and for the Meaney, Merodio--Ogden, Humphrey--Yin, and Gasser--Ogden--Holzapfel models in the anisotropic setting. 
These results indicate that exact representability is not strictly required for useful model discovery. 
However, they also show that the quality of the discovered approximation depends on the expressiveness of the chosen CANN architecture and on the deformation states sampled during training.

A central finding of this study is that constitutive recovery and constitutive generalization must be evaluated separately. 
The CANN architecture used here retains important physical structure, including objectivity, material symmetry, zero energy and zero stress in the reference configuration, and polyconvexity for the selected pseudo-invariants and activation functions. 
Nevertheless, these physical constraints do not by themselves guarantee accurate extrapolation outside the deformation states represented in the training data. 
This distinction was clear in the out-of-function-space benchmarks. 
Acceptable generalization was obtained for the Isihara, Arruda--Boyce, Meaney, and Merodio--Ogden models, whereas poorer generalization was observed for the Gent--Thomas, Ogden, Humphrey--Yin, and Gasser--Ogden--Holzapfel models. 
Even within a single model, generalization was not uniform: for several benchmarks, including Gent--Thomas, Ogden, Meaney, and Merodio--Ogden, the discovered laws performed well in some unseen regions of deformation space but poorly in others. 
This highlights the importance of evaluating discovered constitutive laws over multidimensional deformation domains rather than along a small number of canonical loading paths alone.

Generalization was particularly sensitive in the presence of exponential-type terms.
When the ground-truth response contained an exponential contribution compatible with the CANN functions, as in the Demiray and Holzapfel--Gasser--Ogden benchmarks, the corresponding term could be recovered accurately. 
However, even small discrepancies in exponential terms produced large errors outside the sampled deformation domain because of their rapid growth. 
Conversely, when exponential terms were introduced by the discovered approximation but were not part of the true response, extrapolation also deteriorated, as observed most clearly in the anisotropic benchmarks. 
This does not imply that exponential terms should be avoided.
Rather, it shows that exponential terms require careful interpretation, sufficient probing of the stiffening regime during training, and explicit generalization checks over relevant deformation domains.

The results further highlight the importance of the training deformation-state distribution.
The heterogeneous test we considered here generated a much richer set of deformation states than a conventional homogeneous test, but this set was still imbalanced, with many material points clustered around the dominant biaxial-tension-with-2:1-ratio loading direction. 
The discovered models were most accurate in densely sampled regions and less reliable near the boundaries of the observed domain or in sparsely populated regions. 
This suggests that robust constitutive discovery requires not only an expressive and physically constrained model class, but also experimental designs that deliberately populate the deformation regimes in which the model is expected to be used. 
Recent specimen- and topology-optimization approaches for one-shot material identification provide a natural route toward this goal.

From this perspective, the present findings provide a strong motivation for the stress-unsupervised full-field paradigm. 
The challenge of generalization is not unique to unsupervised discovery. 
If anything, it is more restrictive in conventional supervised calibration, where data are typically obtained from a sparse collection of one-dimensional or nominally homogeneous loading paths. 
By contrast, a single heterogeneous full-field experiment can expose the material model to a substantially broader and more application-relevant portion of deformation space. 
Since accurate constitutive predictions cannot generally be expected far outside the observed domain, improving discovery requires expanding and balancing the domain of seen deformations.
This is precisely where unsupervised full-field discovery offers a fundamental advantage over traditional stress-supervised testing.

Independent forward finite element validation provided a more application-relevant perspective on these generalization errors. 
Although uniform sampling of the multidimensional deformation space revealed large errors in some unseen regions, the median stress errors in forward simulations remained low for most benchmarks. 
Larger discrepancies were typically localized near stress concentrations and occurred in deformation states that were less frequently realized by the mechanical boundary value problem. 
Thus, broad deformation-space error maps are essential for diagnosing the limitations of a discovered law, but they may be conservative relative to practical finite element use, where mechanically realized deformation states occupy only a subset of the full admissible domain.

Several extensions follow naturally from this work. 
First, the framework should be evaluated with broader CANN bases, including generalized invariants, logarithmic terms, principal-stretch-based terms, and other features known to be important in rubber-like materials and soft tissues. 
Second, CANN--EUCLID should be coupled to optimized experimental designs that produce broader and more balanced coverage of application-relevant deformation states. 
Third, the framework should be extended toward realistic biological settings with spatially varying fiber architectures. 
More broadly, the long-term opportunity is to move from discovering a single homogeneous constitutive law toward discovering spatially informed, physics-constrained constitutive representations directly from rich experimental fields. 
Such developments would make unsupervised CANN discovery a practical route toward interpretable material laws for complex and intrinsically heterogeneous soft tissues.

\section{Conclusions}
\label{sec:conclusions}

In this work, we introduced CANN--EUCLID, a stress-unsupervised framework for discovering sparse hyperelastic constitutive laws from full-field displacement data and reaction forces. 
By combining constitutive artificial neural networks (CANNs) with the equilibrium-based training strategy of EUCLID, the framework enables the discovery of sparse interpretable constitutive models that are nonlinear in the material parameters from rich heterogeneous strain states without local stress measurements. 
This approach is particularly suitable for soft biological tissues with strain-stiffening behavior that is canonically captured with exponential terms.

Moreover, we show that calibrating material models on diverse strain states is crucial for obtaining accurate behavior over a large region of the admissible deformation space. 
Therefore, full-field stress-unsupervised training frameworks, such as the one presented in this work, is imperative for material model discovery in the community going forward.

\section*{Data availability}
The unsupervised training framework developed in this work will be released as a module in our COMMET codebase \url{https://github.com/COMMET-code} after publication.
In addition to the unsupervised training of CANNs, the module provides support for training of other neural constitutive models including ICNNs \cite{Amos2017_773c,Thakolkaran2022_37f6}, ICKANs \cite{Liu2025,Thakolkaran2025_30c9}, and custom user-defined neural constitutive models.

\clearpage

\appendix

\section{Polyconvexity of chosen CANN architecture}
\label{sec:cann-polyconvexity}
A strain-energy density function $\Psi\fOf{\F}$ is said to be polyconvex if there exists a \textit{convex} function 
$W: \TT_{+}^{3 \times 3} \times \TT_{+} ^{3\times 3} \times \RR_{+} \rightarrow \RR_{0+}$
such that 
\begin{equation}
    \Psi\fOf{\F}= W\fOf{\F, \adj\F, \det\F}\,,
\end{equation}
where $\TT_{+}^{3 \times 3}$ denotes the set of second-order $3\times3$ tensors with positive determinant, $\RR_{+}$ denotes the set of positive real values, $\RR_{0+}$ denotes the set of non-negative real values, and $\adj\bullet$ denotes the adjoint of $\bullet$.
This condition has been shown to be sufficient for guaranteeing a solution to a finite strain elastic boundary value problem with suitable boundary conditions \cite{Ball1976_4d15} while also allowing for the existence of bifurcations which are physically observed in e.g. elastic buckling.
Additionally, this condition can be verified for a given strain-energy function with relative ease in comparison to the weaker sufficient condition of quasiconvexity \cite{morrey1952quasi,Schroder2003_1753}.
For these reasons, the use of polyconvex strain-energy functions has arguably become the standard approach for guaranteeing a solution to a finite strain elastic boundary value problem in the community.

To keep with this trend, we choose the architecture and pseudo-invariants of the CANN such that the resulting discovered models are guaranteed to be polyconvex.
Note that the architectural choices listed in \sect~\ref{sec:cann-architecture} lead to a model consisting of the summation of the following terms:
\begin{align}
    \theta_{1,j,1} \phi_{1,j,1}\brac{\iIi -3}^{j}\,, \label{eq:I1}\\ 
    \theta_{2,j,1} \phi_{2,j,1}\brac{\iIIi ^{3/2} -3 ^{3/2}}^{j}\,, \label{eq:I2}\\ 
    \theta_{3,j,1} \phi_{3,j,1}\brac{J-1} ^{2j}\,, \label{eq:J}\\ 
    \theta_{4,j,1} \phi_{4,j,1}\mac{\iIVi-1} ^{2j}\,, \label{eq:I4} \\
    \theta_{1,j,2} \left[\exp{\phi_{1,j,2}\brac{\iIi -3} ^{j}}-1\right]\,, \label{eq:exp-first}\\ 
    \theta_{2,j,2} \left[\exp{ \phi_{2,j,2}\brac{\iIIi ^{3/2} -3 ^{3/2}}^{j}}-1\right]\,, \\
    \theta_{3,j,2} \left[\exp{ \phi_{3,j,2}\brac{J-1} ^{2j}}-1\right]\,, \\
    \theta_{4,j,2} \left[\exp{ \phi_{4,j,2}\mac{\iIVi-1} ^{2j}}-1\right]\,. \label{eq:exp-last}
\end{align}
Hence, the discovered model is guaranteed to be polyconvex if $\theta_{ijk},\, \phi_{ijk} \geq 0$ and $j\in\mathbb{N}$, where $\mathbb{N}$ is the set of natural numbers, for the following reasons:
\begin{enumerate}[(i)]
    \item \eqs~\eqref{eq:I1} and \eqref{eq:I2} are proven to be polyconvex  (see Lemma C.6 in \cite{Schroder2003_1753} \footnote{More generally, \smallmath{\brac{\iIi^{k} -3 ^{k}}^{i}} and  \smallmath{\brac{\iIIi^{3k/2} -3^{3k/2} }^{i}} are polyconvex and have zero stress at $\F=\I$ for $i,k \geq 1$, $i,k \in \RR$ \cite{Schroder2003_1753}. Note that $\frac{\norm{\F}^{2}}{\det\F ^{2/3}} = \iIi$ and $ \frac{\norm{\adj\F} ^{3}}{\det \F ^{2}} = \iIIi^{3/2} $ as written in \cite{Schroder2003_1753} Lemma C.6.}). 
    \item \eq~\eqref{eq:J} is convex in $\det\F$ for $j \in \mathbb{N}$. Hence, it is also polyconvex.\footnote{More generally, \smallmath{\brac{J-1} ^{i}} is polyconvex and has zero stress at $\F=\I$ for $i>1$, $i\in\RR$.}
    \item It has been shown that $\iIVi$ is polyconvex (see Lemma C.3, 2 in \cite{Schroder2003_1753}\footnote{Note that $\frac{\tr{\F ^{T}\F \A \otimes \A}^{m}}{\det{\F ^{T}\F} ^{1/3}} = \iIVi$ when $m=1$ as written in \cite{Schroder2003_1753} Lemma C.3.}).
        Additionally, the function $h\fOf{\bullet} = \mac{\bullet - 1}^{2j}$ is convex and monotonically non-decreasing in $\bullet$ for $j\in \mathbb{N}$. \footnote{More generally, \smallmath{\mac{\iIVi-1} ^{p}} is polyconvex and has zero stress at $\F=\I$ for $p > 1$ , $p \in \RR$ \cite{Balzani2006_50ce}.}
        Hence, \eq~\eqref{eq:I4} is polyconvex. 
    \item Since $\exp\bullet$ is convex and monotonically increasing in $\bullet$ and \eqs~\eqref{eq:I1}--\eqref{eq:I4} are polyconvex, it follows that \eqs~\eqref{eq:exp-first}--\eqref{eq:exp-last} are also polyconvex. 
    \item The summation of polyconvex functions is polyconvex.
\end{enumerate}

\section{Training details}
\label{sec:training-details}

\subsection{Avoiding constrained optimization}
\label{sec:training-details-avoidconstraintopt}

As discussed in \ref{sec:cann-polyconvexity}, the CANN architecture used in this work is guaranteed to be polyconvex provided that the weights satisfy
\begin{equation}
    \theta_{ijk}\geq 0,
    \qquad
    \phi_{ijk}\geq 0.
\end{equation}
These non-negativity conditions could be imposed directly during training \cite{Linka2023_6533}, but doing so would turn the problem into a constrained optimization problem.
This is inconvenient in the present setting, where we use standard gradient-based neural-network optimizers such as Adam and where the loss is already complicated by nonlinearity and sparsity-promoting regularization.
We therefore enforce non-negativity by reparameterizing the physical CANN weights in terms of unconstrained auxiliary variables.
Following the strategy used in \cite{As_ad2022_6b13}, we introduce unconstrained auxiliary parameters $\hat{\theta}_{ijk},\hat{\phi}_{ijk}\in\RR$, and define the physical non-negative weights through the softplus function $s$,
\begin{align}
    \theta_{ijk}
    &=
    s\fOf{\hat{\theta}_{ijk}},
    &
    \phi_{ijk}
    &=
    s\fOf{\hat{\phi}_{ijk}},
    &
    s\fOf{\bullet}
    &=
    \log\fOf{1+\exp{\bullet}}.
    \label{eq:softplus-reparameterization}
\end{align}
The softplus function maps real-valued inputs to strictly positive outputs, i.e. $s:\RR\rightarrow\RR_{+}$, and is smooth and monotonically increasing. 
Consequently, the auxiliary parameters $\hat{\theta}_{ijk}$ and $\hat{\phi}_{ijk}$ can be optimized freely over the entire real line, while the corresponding CANN weights $\theta_{ijk}$ and $\phi_{ijk}$ remain non-negative by construction. 
The original constrained problem in the physical weights is thus replaced by an unconstrained problem in the auxiliary parameters.

In implementation, the learnable parameters are $\hat{\theta}_{ijk}$ and $\hat{\phi}_{ijk}$, while all evaluations of the strain-energy density, stresses, equilibrium residuals, and regularization terms use the transformed weights $\theta_{ijk}$ and $\phi_{ijk}$ from \eq~\eqref{eq:softplus-reparameterization}. 
Thus, the CANN strain-energy density is evaluated as
\begin{equation}
    \en
    =
    \sum_{i}^{n_{\actK}}
    \sum_{j}^{n_f}
    \sum_{k}^{n_g}
    s\fOf{\hat{\theta}_{ijk}}
    g_k\fOf{
        s\fOf{\hat{\phi}_{ijk}}
        \actK_i^j
    }\,.
    \label{eq:cann-softplus}
\end{equation}

\subsection{Data normalization}
\label{sec:data-normalization}

It is well known in the machine-learning community that the success and robustness of neural-network training depend strongly on the normalization of the data \cite{LeCun1998_77c8,Huang2023_681f}. 
In constitutive model discovery, however, normalization must be applied with care. 
Standard preprocessing strategies, such as subtracting the mean of each input feature and scaling by its standard deviation, are not directly suitable for hyperelastic neural constitutive models. 
For example, translating the pseudo-invariants would generally destroy the property that all inputs vanish in the reference configuration, which may lead to a non-zero strain-energy density or a non-zero stress at $\F=\I$. 
We therefore use normalization strategies that improve numerical conditioning while preserving the physical and mathematical structure of the CANN.

\subsubsection*{Input normalization}

The purpose of input normalization is to prevent different pseudo-invariant terms from entering the optimization problem at widely different scales. 
If, for example, one pseudo-invariant varies over a range of order $10^{0}$ while another varies over a range of order $10^{3}$, the gradients associated with the latter can dominate the optimization. 
This imbalance leads to an ill-conditioned loss landscape and can make training sensitive to initialization and learning-rate choices. 
For CANNs, this issue is particularly important because some terms contain exponential activation functions. 
Large unnormalized inputs to the exponential terms can cause numerical overflow or strongly bias the optimizer toward suppressing these terms before they have been meaningfully calibrated.

At the same time, the normalization must preserve the defining properties of the pseudo-invariants discussed in \sect~\ref{sec:pseudo-invariants}. 
In particular, it should not shift the inputs away from zero at the reference configuration and should not invalidate the polyconvexity construction. 
For this reason, we only rescale the pseudo-invariant powers by positive constants. 
Specifically, instead of using the unnormalized CANN
\begin{equation}
    \en
    =
    \sum_{i}^{n_{\actK}}
    \sum_{j}^{n_f}
    \sum_{k}^{n_g}
    \theta_{ijk}
    g_k\fOf{
        \phi_{ijk}
        \actK_i^j
    }\,,
\end{equation}
we use the normalized form
\begin{equation}
    \en
    =
    \sum_{i}^{n_{\actK}}
    \sum_{j}^{n_f}
    \sum_{k}^{n_g}
    \theta_{ijk}
    g_k\fOf{
        \phi_{ijk}
        c_{ij}
        \actK_i^j
    }\,,
    \label{eq:input-normalized-cann}
\end{equation}
where the constants $c_{ij}$ are chosen such that
\begin{equation}
    \text{(no sum)} \qquad c_{ij} \abs{\actK_i^j} \leq 1\,,
\end{equation}
for the given dataset.
This is done by converting the dataset of nodal displacement values $\uSet$ to a dataset of deformation gradient values at quadrature points $\FDSet$ through \eq~\eqref{eq:f-calc}, and obtaining $c_{ij}$ using
\begin{equation}
    c_{ij}
    =
    \frac{1}{
    \max_{\F\in\FDSet}
    \abs{
    \actK_i^j
    }
    } \,.
    \label{eq:input-normalization-constant}
\end{equation}
This normalization maps the largest observed magnitude of each pseudo-invariant power to one. 
Importantly, because the normalization consists only of multiplication by positive constants, the normalized pseudo-invariant powers still vanish in the reference configuration. 
Moreover, the zero-stress arguments from \sect~\ref{sec:pseudo-invariants} remain unchanged, and the polyconvexity arguments in \ref{sec:cann-polyconvexity} continue to hold for non-negative trainable weights. 
Thus, \eq~\eqref{eq:input-normalized-cann} improves numerical conditioning without changing the physical admissibility of the CANN architecture.

In the absence of input normalization one may formally set $c_{ij}=1$ for all branches. 
In practice, however, we found the input normalization in \eq~\eqref{eq:input-normalization-constant} to substantially improve training robustness, especially for models containing exponential terms.

\subsubsection*{Output normalization}

Output normalization is also important for robust CANN training, particularly when sparsity-promoting $L_p$ regularization is used \cite{McCulloch2024_6c1e}. 
In stress-supervised training, output normalization is straightforward because the target stresses are known and can be scaled directly. 
In the stress-unsupervised setting considered here, local stress targets are unavailable. 
Nevertheless, the reaction forces provide an experimentally accessible measure of the overall force scale and can therefore be used to normalize the contributions of internal and external force balance to the loss.

We define the characteristic reaction-force magnitude
\begin{equation}
    R_0
    :=
    \max_{i,t,\beta}
    \abs{R_i^{\beta,t}},
    \label{eq:reaction-force-scale}
\end{equation}
and introduce normalized reaction forces
\begin{equation}
    \hat{R}_i^{\beta,t}
    :=
    \frac{R_i^{\beta,t}}{R_0}.
    \label{eq:normalized-reaction-force}
\end{equation}
During training, the CANN is interpreted as representing a normalized strain-energy density
\begin{equation}
    \hat{\en}\fOf{\F;\params}
    :=
    \frac{1}{R_0}
    \en\fOf{\F;\params}.
    \label{eq:normalized-energy}
\end{equation}
Consequently, the corresponding first Piola--Kirchhoff stress and internal nodal forces are also normalized:
\begin{equation}
    \hat{\P}
    =
    \pd{\hat{\en}}{\F}
    =
    \frac{1}{R_0}\P,
    \qquad
    \hat{r}_{\inter,i}^{I,t}
    =
    \intOnDomO{
    \hat{P}_{ij}\fOf{\F^t,\aSet;\params}
    \Grad{\shapeFnI}_j
    }
    =
    \frac{1}{R_0}
    r_{\inter,i}^{I,t}.
    \label{eq:normalized-internal-force}
\end{equation}
The equilibrium terms in the loss are then evaluated using these normalized quantities:
\begin{align}
    \hat{L}_{\inter}
    &=
    \frac{1}{\nSteps\nNodes}
    \sum_t
    \sum_{(I,i)\in D^{\text{free}}}
    \brac{
    \hat{r}_{\inter,i}^{I,t}
    }^2,
    \label{eq:normalized-l-free}
    \\
    \hat{L}_{\ext}
    &=
    \frac{1}{\nSteps\nBeta}
    \sum_{\beta,t}
    \brac{
    \hat{\R}^{\beta,t}
    -
    \sum_{I\in D^\beta}
    \hat{\r}_{\inter}^{I,t}
    }^2.
    \label{eq:normalized-l-fixed}
\end{align}
Thus, the loss, when output normalization is taken into account, takes the form
\begin{equation}
    L =
    \lambda_{\inter}\hat{L}_{\inter}
    +
    \lambda_{\ext}\hat{L}_{\ext}
    +
    \lambda_p L_p\,,
    \label{eq:normalized-loss}
\end{equation}
where $L_p$ is given by \eq~\eqref{eq:our-reg}.

This normalization makes the magnitude of the equilibrium residual largely independent of the absolute stiffness of the ground-truth material. 
As a result, the same weight initialization, learning rates, and regularization weights can be used across materials with substantially different stress scales. 
After training, the physical strain-energy density is recovered by reversing the scaling,
\begin{equation}
    \en\fOf{\F;\params}
    =
    R_0
    \hat{\en}\fOf{\F;\params},
    \label{eq:recover-physical-energy}
\end{equation}
and the physical stress follows accordingly as
\begin{equation}
    \P\fOf{\F;\params}
    =
    R_0
    \pd{\hat{\en}\fOf{\F;\params}}{\F}.
    \label{eq:recover-physical-stress}
\end{equation}

\subsection{Initialization}

Because output normalization scales the reaction forces such that their maximum magnitude is unity, the normalized internal forces generated by the initial CANN should also be of order one. 
Similarly, because input normalization scales each pseudo-invariant power such that its maximum magnitude over the training set is one, the arguments of the activation functions can be controlled directly through the inner weights $\phi_{ijk}$. 
We therefore initialize the weights so that
\begin{equation}
    \phi_{ijk}=1,
    \qquad
    \sum_{i}^{n_{\actK}}
    \sum_{j}^{n_f}
    \sum_{k}^{n_g}
    \theta_{ijk}
    =
    1.
    \label{eq:init-physical-weights}
\end{equation}
The choice $\phi_{ijk}=1$ gives each branch an initially comparable sensitivity to its normalized pseudo-invariant input. 
The normalization of the outer weights $\theta_{ijk}$ ensures that the total initial magnitude of the CANN response remains approximately independent of the number of candidate terms in the architecture. 
In other words, adding more pseudo-invariants, powers, or activation functions does not automatically increase the initial stress scale.

To avoid initializing all terms identically, the initial values of the outer weights are obtained by applying a softmax transform to a vector of samples from a normal distribution, i.e. 
\begin{align}
    z_{ijk}&\sim\mathcal{N}\fOf{0,\sigma_{\text{init}}^2},&
    \theta_{ijk}
           &=
    \frac{\exp{z_{ijk}}}
    {
    \sum_{a}^{n_{\actK}}
    \sum_{b}^{n_f}
    \sum_{c}^{n_g}
    \exp{z_{abc}}
    }\,,
    \label{eq:theta-softmax-init}
\end{align}
where the variance $\sigma_{\text{init}}^2$ is a hyperparameter that controls the degree of initial variation between terms: small values produce nearly uniform initial weights, whereas larger values give some terms slightly larger initial contributions. 
This construction guarantees that all initial $\theta_{ijk}$ are positive and that their sum is equal to one. 

As described in \sect~\ref{sec:training-details-avoidconstraintopt}, the actual learnable parameters are not the weights $\theta_{ijk}$ and $\phi_{ijk}$, but the unconstrained auxiliary parameters $\hat{\theta}_{ijk}$ and $\hat{\phi}_{ijk}$, which are mapped to the physical weights through the softplus function,
After selecting the desired initial physical weights from \eqs~\eqref{eq:init-physical-weights} and \eqref{eq:theta-softmax-init}, the corresponding auxiliary parameters are obtained by applying the inverse softplus map,
\begin{align}
    \hat{\theta}_{ijk}
    &=
    s^{-1}\fOf{\theta_{ijk}},
    &
    \hat{\phi}_{ijk}
    &=
    s^{-1}\fOf{\phi_{ijk}},
    &
    s^{-1}\fOf{\bullet}
    &=
    \log\fOf{\exp{\bullet}-1}.
    \label{eq:inverse-softplus-init}
\end{align}

This initialization strategy is tied to the normalization procedure used in this work. 
Without input normalization, setting $\phi_{ijk}=1$ could lead to very large activation-function arguments if some pseudo-invariant powers have large magnitudes. 
Similarly, without output normalization, the appropriate magnitude of the outer weights $\theta_{ijk}$ would depend on the unknown stiffness scale of the material. 
Thus, the initialization described above should be interpreted as part of the combined normalization--initialization strategy: input normalization controls the scale of the CANN inputs, output normalization controls the scale of the equilibrium residuals, and the initialization controls the initial magnitude of the CANN stress predictions.

\subsection{Multistage training}

Despite the utility of $L_p$ regularization, it also introduces some challenges.
Such challenges include:
\begin{enumerate}[(i)]
    \item The final selection of terms being highly sensitive to model initialization \cite{McCulloch2024_6c1e}. 
    \item The \textit{optimized} parameters being suboptimal for the primary objective (i.e. minimizing the imbalance of equilibrium) due to the fact that the $L_p$-norm must be minimized simultaneously.
\end{enumerate}
To address these challenges, we take a three-stage training approach:

\textit{(Stage 1) Pre-regularization.}
\begin{adjustwidth}{12pt}{}
    The sensitivity to model initialization arises from the fact that terms closer to zero are penalized more strongly by $L_p$ regularization than terms further from zero.
    Consequently, terms close to zero by happenstance after initialization are arbitrarily penalized more than others.
    To reduce this sensitivity, we first train without imposing $L_p$ regularization, i.e. we set $\lambda_p=0$.
    Doing so allows each term to naturally adjust its contribution to the overall behavior from its initialized state before any term is penalized.
\end{adjustwidth}
\textit{(Stage 2) Regularization and subset selection.}
\begin{adjustwidth}{12pt}{}
    We then apply $L_p$ regularization to induce sparsity for a chosen number of epochs.
    At the end of this stage, we select all terms that have a non-negligible contribution to the overall behavior.
    This is determined by evaluating the strain-energy density for each deformation gradient in the dataset and keeping track of the proportion that is contributed by each term.
    Terms that have an average proportional contribution above a small chosen cut-off threshold are selected and all other terms are set to zero.
\end{adjustwidth}
\textit{(Stage 3) Post-regularization polishing.}
\begin{adjustwidth}{12pt}{}
    After selecting the subset of active terms, we again train without imposing $L_p$ regularization, i.e. we set $\lambda_p=0$ again.
    Hence, the parameters of terms in the resulting subset are truly optimized for minimizing the imbalance of equilibrium.
\end{adjustwidth}
Taken together, these three stages reduce sensitivity to model initialization and ensure that the parameter values of the selected model truly minimize the imbalance of equilibrium.

\subsection{Datasets and hyperparameters}
\label{sec:hyperparameters}
Further details and values for the training and validation data generation along with model and training hyperparameters are presented in \tab~\ref{tab:parameters}.

\begin{table}[tb]
\centering
\caption{Parameters and hyperparameters used for the data generation, validation, and model training.} 
\label{tab:parameters}
\begin{tabular}{lcc}
\hline
\multicolumn{1}{l}{\textbf{Parameter}} & \textbf{Notation} & \textbf{Value} \\ 
\hline
\textit{Training specimen:}\\
$\quad$ Number of nodes   & $n_n$ & $1\,464$  \\
$\quad$ Number of reaction force constraints & $n_{\beta}$   & $6$   \\ 
$\quad$ Number of data snapshots     & $n_{t}$     & 10 \\
$\quad$ Loading parameter   & $\delta$  & $\{0.1 \times t: t=1,\dots,n_t\}$ \\
\textit{Validation specimen:}\\
$\quad$ Number of nodes  & - & $20\,082$ \\
\hline
\textit{CANN hyperparameters:}\\
$\quad$ Number of power functions & $n_f$    & $2$\\
$\quad$ Number of activation functions & $n_g$    & $2$\\
$\quad$ Activation functions & $-$    & $\bullet$, $\exp{\bullet} - 1$\\
$\quad$ Pseudo-invariants (isotropic cases) & $-$    &$\iIi -3$, $\iIIi^{3/2} - 3 ^{3/2}$, $ \brac{J-1}^{2}$ \\
$\quad$ Pseudo-invariants (anisotropic cases) & $-$    &$\iIi -3$, $\iIIi^{3/2} - 3 ^{3/2}$, $ \brac{J-1}^{2}$, $\mac{\iIVi - 1}^{2}$\\
\hline
\textit{Training hyperparameters:}\\
$\quad$ \textit{(Stage 1) Pre-regularization:}\\
$\quad\quad$ Internal force balance weight  & $\lambda_\inter$    & 1  \\
$\quad\quad$ External force balance weight  & $\lambda_\ext$    & 1  \\
$\quad\quad$ Regularization weight  & $\lambda_p$    & 0  \\
$\quad\quad$ Optimizer  & $-$    & Adam  \\
$\quad\quad$ Epochs  & $-$    & 4000  \\
$\quad\quad$ Learning rate  & $-$    & 0.025 \\
$\quad$ \textit{(Stage 2) Regularization and subset selection:}\\
$\quad\quad$ Internal force balance weight  & $\lambda_\inter$    & 1  \\
$\quad\quad$ External force balance weight  & $\lambda_\ext$    & 1  \\
$\quad\quad$ Regularization weight  & $\lambda_p$    & 0.001  \\
$\quad\quad$ Regularization exponent  & $p$    & 0.25  \\
$\quad\quad$ Optimizer  & $-$    & Adam  \\
$\quad\quad$ Epochs  & $-$    & 4000  \\
$\quad\quad$ Learning rate  & $-$    & 0.025 \\
$\quad\quad$ Subset cut-off threshold  & $-$    & $1\times 10 ^{-4}$ \\
$\quad$ \textit{(Stage 3) Post-regularization polishing:} \\
$\quad\quad$ Internal force balance weight  & $\lambda_\inter$    & 1  \\
$\quad\quad$ External force balance weight  & $\lambda_\ext$    & 1  \\
$\quad\quad$ Regularization weight  & $\lambda_p$    & 0  \\
$\quad\quad$ Optimizer  & $-$    & Adam  \\
$\quad\quad$ Epochs  & $-$    & 4000  \\
$\quad\quad$ Learning rate  & $-$    & 0.005 \\
\hline
\end{tabular}
\end{table}

\clearpage

\bibliographystyle{elsarticle-num}
\bibliography{bib-file}

\begin{thebibliography}{100}
\expandafter\ifx\csname url\endcsname\relax
  \def\url#1{\texttt{#1}}\fi
\expandafter\ifx\csname urlprefix\endcsname\relax\def\urlprefix{URL }\fi
\expandafter\ifx\csname href\endcsname\relax
  \def\href#1#2{#2} \def\path#1{#1}\fi

\bibitem{Ghaboussi_576d}
J.~Ghaboussi, J.~H. Garrett, X.~Wu, \href{https://ascelibrary.org/doi/10.1061/%28ASCE%290733-9399%281991%29117%3A1%28132%29}{Knowledge‐based modeling of material behavior with neural networks}, Journal of Engineering Mechanics 117  132--153.
\newblock \href {https://doi.org/10.1061/(ASCE)0733-9399(1991)117:1(132)} {\path{doi:10.1061/(ASCE)0733-9399(1991)117:1(132)}}.
\newline\urlprefix\url{https://ascelibrary.org/doi/10.1061/%28ASCE%290733-9399%281991%29117%3A1%28132%29}

\bibitem{Masi2021_3239}
F.~Masi, I.~Stefanou, P.~Vannucci, V.~Maffi-Berthier, \href{https://doi.org/10.1016/j.jmps.2020.104277}{Thermodynamics-based artificial neural networks for constitutive modeling}, Journal of the Mechanics and Physics of Solids 147 (2021) 104277.
\newblock \href {https://doi.org/10.1016/j.jmps.2020.104277} {\path{doi:10.1016/j.jmps.2020.104277}}.
\newline\urlprefix\url{https://doi.org/10.1016/j.jmps.2020.104277}

\bibitem{Vlassis2021_7f8b}
N.~N. Vlassis, W.~Sun, \href{https://doi.org/10.1016/j.cma.2021.113695}{Sobolev training of thermodynamic-informed neural networks for interpretable elasto-plasticity models with level set hardening}, Computer Methods in Applied Mechanics and Engineering 377 (2021) 113695.
\newblock \href {https://doi.org/10.1016/j.cma.2021.113695} {\path{doi:10.1016/j.cma.2021.113695}}.
\newline\urlprefix\url{https://doi.org/10.1016/j.cma.2021.113695}

\bibitem{Tac2022_a940}
V.~Tac, V.~D. Sree, M.~K. Rausch, A.~B. Tepole, \href{http://dx.doi.org/10.1007/s00366-022-01733-3}{Data-driven modeling of the mechanical behavior of anisotropic soft biological tissue}, Engineering with Computers 38 (2022) 4167--4182.
\newblock \href {https://doi.org/10.1007/s00366-022-01733-3} {\path{doi:10.1007/s00366-022-01733-3}}.
\newline\urlprefix\url{http://dx.doi.org/10.1007/s00366-022-01733-3}

\bibitem{Tac2022_7978}
V.~Tac, F.~Sahli~Costabal, A.~B. Tepole, \href{http://dx.doi.org/10.1016/j.cma.2022.115248}{Data-driven tissue mechanics with polyconvex neural ordinary differential equations}, Computer Methods in Applied Mechanics and Engineering 398 (2022) 115248.
\newblock \href {https://doi.org/10.1016/j.cma.2022.115248} {\path{doi:10.1016/j.cma.2022.115248}}.
\newline\urlprefix\url{http://dx.doi.org/10.1016/j.cma.2022.115248}

\bibitem{Linden2023_590d}
L.~Linden, D.~K. Klein, K.~A. Kalina, J.~Brummund, O.~Weeger, M.~K{\"a}stner, \href{http://dx.doi.org/10.1016/j.jmps.2023.105363}{Neural networks meet hyperelasticity: A guide to enforcing physics}, Journal of the Mechanics and Physics of Solids 179 (2023) 105363.
\newblock \href {https://doi.org/10.1016/j.jmps.2023.105363} {\path{doi:10.1016/j.jmps.2023.105363}}.
\newline\urlprefix\url{http://dx.doi.org/10.1016/j.jmps.2023.105363}

\bibitem{Dornheim2024_5d67}
J.~Dornheim, L.~Morand, H.~J. Nallani, D.~Helm, \href{https://doi.org/10.1007/s11831-023-10009-y}{Neural networks for constitutive modeling: From universal function approximators to advanced models and the integration of physics}, Archives of Computational Methods in Engineering 31 (2024) 1097--1127.
\newblock \href {https://doi.org/10.1007/s11831-023-10009-y} {\path{doi:10.1007/s11831-023-10009-y}}.
\newline\urlprefix\url{https://doi.org/10.1007/s11831-023-10009-y}

\bibitem{Fuhg2024_7694}
J.~N. Fuhg, G.~Anantha~Padmanabha, N.~Bouklas, B.~Bahmani, W.~Sun, N.~N. Vlassis, M.~Flaschel, P.~Carrara, L.~De~Lorenzis, A review on data-driven constitutive laws for solids, Archives of Computational Methods in Engineering (2024) 1--43.

\bibitem{Tac2024_5fa1}
V.~Tac, E.~Kuhl, A.~B. Tepole, \href{https://doi.org/10.1016/j.eml.2024.102220}{Data-driven continuum damage mechanics with built-in physics}, Extreme Mechanics Letters 71 (2024) 102220.
\newblock \href {https://doi.org/10.1016/j.eml.2024.102220} {\path{doi:10.1016/j.eml.2024.102220}}.
\newline\urlprefix\url{https://doi.org/10.1016/j.eml.2024.102220}

\bibitem{Tac2025_1366}
V.~Ta{\c c}, M.~K. Rausch, I.~Bilionis, F.~Sahli~Costabal, A.~B. Tepole, \href{https://doi.org/10.1007/s00366-024-01984-2}{Generative hyperelasticity with physics-informed probabilistic diffusion fields}, Engineering with Computers 41 (2025) 51--69.
\newblock \href {https://doi.org/10.1007/s00366-024-01984-2} {\path{doi:10.1007/s00366-024-01984-2}}.
\newline\urlprefix\url{https://doi.org/10.1007/s00366-024-01984-2}

\bibitem{Alheit2026_32d4}
B.~Alheit, M.~Peirlinck, S.~Kumar, \href{https://doi.org/10.1016/j.cma.2026.118728}{Commet: Orders-of-magnitude speed-up in finite element method via batch-vectorized neural constitutive updates}, Computer Methods in Applied Mechanics and Engineering 452 (2026) 118728.
\newblock \href {https://doi.org/10.1016/j.cma.2026.118728} {\path{doi:10.1016/j.cma.2026.118728}}.
\newline\urlprefix\url{https://doi.org/10.1016/j.cma.2026.118728}

\bibitem{Jadoon2025_2f3f}
A.~A. Jadoon, K.~A. Kalina, M.~K. Rausch, R.~Jones, J.~N. Fu~hg, \href{https://doi.org/10.1016/j.jmps.2025.106161}{Inverse design of anisotropic microstructures using physics-augmented neural networks}, Journal of the Mechanics and Physics of Solids 203 (2025) 106161.
\newblock \href {https://doi.org/10.1016/j.jmps.2025.106161} {\path{doi:10.1016/j.jmps.2025.106161}}.
\newline\urlprefix\url{https://doi.org/10.1016/j.jmps.2025.106161}

\bibitem{Flaschel2026_762c}
M.~Flaschel, D.~Martonov{\'a}, C.~Veil, E.~Kuhl, \href{https://doi.org/10.1016/j.cma.2025.118573}{Material fingerprinting: A shortcut to material model discovery without solving optimization problems}, Computer Methods in Applied Mechanics and Engineering 450 (2026) 118573.
\newblock \href {https://doi.org/10.1016/j.cma.2025.118573} {\path{doi:10.1016/j.cma.2025.118573}}.
\newline\urlprefix\url{https://doi.org/10.1016/j.cma.2025.118573}

\bibitem{Martonova2026_20bb}
D.~Martonov{\'a}, E.~Kuhl, M.~Flaschel, \href{https://doi.org/10.1016/j.jmps.2025.106463}{Material fingerprinting for rapid discovery of hyperelastic models: First experimental validation}, Journal of the Mechanics and Physics of Solids 208 (2026) 106463.
\newblock \href {https://doi.org/10.1016/j.jmps.2025.106463} {\path{doi:10.1016/j.jmps.2025.106463}}.
\newline\urlprefix\url{https://doi.org/10.1016/j.jmps.2025.106463}

\bibitem{Flaschel2026_3b56}
M.~Flaschel, M.~A. Moreno-Mateos, S.~Wiesheier, P.~Steinmann, E.~Kuhl, \href{http://arxiv.org/abs/2601.14965}{Unsupervised material fingerprinting: Ultra-fast hyperelastic model discovery from full-field experimental measurements} (1 2026).
\newblock \href {http://arxiv.org/abs/2601.14965} {\path{arXiv:2601.14965}}, \href {https://doi.org/10.48550/arXiv.2601.14965} {\path{doi:10.48550/arXiv.2601.14965}}.
\newline\urlprefix\url{http://arxiv.org/abs/2601.14965}

\bibitem{Klein2022_3243}
D.~K. Klein, M.~Fern{\'a}ndez, R.~J. Martin, P.~Neff, O.~Weeger, \href{https://doi.org/10.1016/j.jmps.2021.104703}{Polyconvex anisotropic hyperelasticity with neural networks}, Journal of the Mechanics and Physics of Solids 159 (2022) 104703.
\newblock \href {https://doi.org/10.1016/j.jmps.2021.104703} {\path{doi:10.1016/j.jmps.2021.104703}}.
\newline\urlprefix\url{https://doi.org/10.1016/j.jmps.2021.104703}

\bibitem{Magana2025_291c}
E.~Maga{\~n}a, S.~Pezzuto, F.~Sahli~Costabal, \href{https://doi.org/10.1113/jp288001}{Ensemble learning of the atrial fibre orientation with physics-informed neural networks}, The Journal of Physiology n/a (9 2025).
\newblock \href {https://doi.org/10.1113/jp288001} {\path{doi:10.1113/jp288001}}.
\newline\urlprefix\url{https://doi.org/10.1113/jp288001}

\bibitem{Flaschel2023_773f}
M.~Flaschel, H.~Yu, N.~Reiter, J.~Hinrichsen, S.~Budday, P.~Steinmann, S.~Kumar, L.~De~Lorenzis, \href{https://doi.org/10.1016/j.jmps.2023.105404}{Automated discovery of interpretable hyperelastic material models for human brain tissue with euclid}, Journal of the Mechanics and Physics of Solids 180 (2023) 105404.
\newblock \href {https://doi.org/10.1016/j.jmps.2023.105404} {\path{doi:10.1016/j.jmps.2023.105404}}.
\newline\urlprefix\url{https://doi.org/10.1016/j.jmps.2023.105404}

\bibitem{Urrea_Quintero2026_474e}
J.~H. Urrea\textendash~Quintero, D.~Anton, L.~De~Lorenzis, H.~Wessels, \href{https://doi.org/10.1016/j.cma.2025.118551}{Automated constitutive model discovery by pairing sparse regression algorithms with model selection criteria}, Computer Methods in Applied Mechanics and Engineering 449 (2026) 118551.
\newblock \href {https://doi.org/10.1016/j.cma.2025.118551} {\path{doi:10.1016/j.cma.2025.118551}}.
\newline\urlprefix\url{https://doi.org/10.1016/j.cma.2025.118551}

\bibitem{Schoenauer1996_725e}
M.~Schoenauer, M.~Sebag, F.~Jouve, B.~Lamy, H.~Maitournam, \href{https://hal.science/hal-00112301}{Evolutionary identification of macro-mechanical models}, MIT Press, 1996, pp. 467--488.
\newline\urlprefix\url{https://hal.science/hal-00112301}

\bibitem{Wang2022_49fb}
M.~Wang, C.~Chen, W.~Liu, \href{https://doi.org/10.1016/j.jmps.2021.104742}{Establish algebraic data-driven constitutive models for elastic solids with a tensorial sparse symbolic regression method and a hybrid feature selection technique}, Journal of the Mechanics and Physics of Solids 159 (2022) 104742.
\newblock \href {https://doi.org/10.1016/j.jmps.2021.104742} {\path{doi:10.1016/j.jmps.2021.104742}}.
\newline\urlprefix\url{https://doi.org/10.1016/j.jmps.2021.104742}

\bibitem{Abdusalamov2023_6244}
R.~Abdusalamov, M.~Hillg{\"a}rtner, M.~Itskov, \href{https://doi.org/10.1002/nme.7203}{Automatic generation of interpretable hyperelastic material models by symbolic regression}, International Journal for Numerical Methods in Engineering 124 (2023) 2093--2104.
\newblock \href {https://doi.org/10.1002/nme.7203} {\path{doi:10.1002/nme.7203}}.
\newline\urlprefix\url{https://doi.org/10.1002/nme.7203}

\bibitem{Hou2024_1ba4}
J.~Hou, X.~Chen, T.~Wu, E.~Kuhl, X.~Wang, \href{https://doi.org/10.1016/j.actbio.2024.09.005}{Automated data-driven discovery of material models based on symbolic regression: A case study on the human brain cortex}, Acta Biomaterialia 188 (2024) 276--296.
\newblock \href {https://doi.org/10.1016/j.actbio.2024.09.005} {\path{doi:10.1016/j.actbio.2024.09.005}}.
\newline\urlprefix\url{https://doi.org/10.1016/j.actbio.2024.09.005}

\bibitem{Kusner2017_7f94}
M.~J. Kusner, B.~Paige, J.~M. Hern{\'a}ndez-Lobato, \href{https://proceedings.mlr.press/v70/kusner17a.html}{Grammar variational autoencoder}, in: Proceedings of the 34th International Conference on Machine Learning, PMLR, 2017, pp. 1945--1954.
\newline\urlprefix\url{https://proceedings.mlr.press/v70/kusner17a.html}

\bibitem{Kissas2024_1676}
G.~Kissas, S.~Mishra, E.~Chatzi, L.~De~Lorenzis, \href{https://doi.org/10.1016/j.cma.2024.117053}{The language of hyperelastic materials}, Computer Methods in Applied Mechanics and Engineering 428 (2024) 117053.
\newblock \href {https://doi.org/10.1016/j.cma.2024.117053} {\path{doi:10.1016/j.cma.2024.117053}}.
\newline\urlprefix\url{https://doi.org/10.1016/j.cma.2024.117053}

\bibitem{Liu2025}
Z.~Liu, Y.~Wang, S.~Vaidya, F.~Ruehle, J.~Halverson, M.~Solja{\v c}i{\' c}, T.~Y. Hou, M.~Tegmark, \href{http://arxiv.org/abs/2404.19756}{Kan: Kolmogorov-arnold networks} (2 2025).
\newblock \href {http://arxiv.org/abs/2404.19756} {\path{arXiv:2404.19756}}, \href {https://doi.org/10.48550/arXiv.2404.19756} {\path{doi:10.48550/arXiv.2404.19756}}.
\newline\urlprefix\url{http://arxiv.org/abs/2404.19756}

\bibitem{Thakolkaran2025_30c9}
P.~Thakolkaran, Y.~Guo, S.~Saini, M.~Peirlinck, B.~Alheit, S.~Kumar, Can kan cans? input-convex kolmogorov-arnold networks (kans) as hyperelastic constitutive artificial neural networks (cans), Computer Methods in Applied Mechanics and Engineering 443 (2025) 118089.

\bibitem{Abdolazizi2025}
K.~P. Abdolazizi, R.~C. Aydin, C.~J. Cyron, K.~Linka, \href{http://dx.doi.org/10.1016/j.jmps.2025.106212}{Constitutive kolmogorov–arnold networks (ckans): Combining accuracy and interpretability in data-driven material modeling}, Journal of the Mechanics and Physics of Solids 203 (2025) 106212.
\newblock \href {https://doi.org/10.1016/j.jmps.2025.106212} {\path{doi:10.1016/j.jmps.2025.106212}}.
\newline\urlprefix\url{http://dx.doi.org/10.1016/j.jmps.2025.106212}

\bibitem{Fuhg2024_55bc}
J.~N. Fuhg, R.~E. Jones, N.~Bouklas, \href{https://doi.org/10.1016/j.cma.2024.116973}{Extreme sparsification of physics-augmented neural networks for interpretable model discovery in mechanics}, Computer Methods in Applied Mechanics and Engineering 426 (2024) 116973.
\newblock \href {https://doi.org/10.1016/j.cma.2024.116973} {\path{doi:10.1016/j.cma.2024.116973}}.
\newline\urlprefix\url{https://doi.org/10.1016/j.cma.2024.116973}

\bibitem{Anantha_Padmanabha2024_35d1}
G.~Anantha~Padmanabha, J.~N. Fuhg, C.~Safta, R.~E. Jones, N.~Bouklas, \href{https://doi.org/10.1016/j.cma.2024.117359}{Improving the performance of stein variational inference through extreme sparsification of physically-constrained neural network models}, Computer Methods in Applied Mechanics and Engineering 432 (2024) 117359.
\newblock \href {https://doi.org/10.1016/j.cma.2024.117359} {\path{doi:10.1016/j.cma.2024.117359}}.
\newline\urlprefix\url{https://doi.org/10.1016/j.cma.2024.117359}

\bibitem{Tan2026_13fd}
J.~Tan, G.~A. Padmanabha, S.~J. Yang, N.~Bouklas, \href{http://arxiv.org/abs/2604.07746}{Towards rapid constitutive model discovery from multi-modal data: Physics augmented finite element model updating (pafemu)} (4 2026).
\newblock \href {http://arxiv.org/abs/2604.07746} {\path{arXiv:2604.07746}}, \href {https://doi.org/10.48550/arXiv.2604.07746} {\path{doi:10.48550/arXiv.2604.07746}}.
\newline\urlprefix\url{http://arxiv.org/abs/2604.07746}

\bibitem{Bahmani2024_4956}
B.~Bahmani, W.~Sun, \href{https://doi.org/10.1002/nme.7473}{Physics‐constrained symbolic model discovery for polyconvex incompressible hyperelastic materials}, International Journal for Numerical Methods in Engineering 125 (2024) e7473.
\newblock \href {https://doi.org/10.1002/nme.7473} {\path{doi:10.1002/nme.7473}}.
\newline\urlprefix\url{https://doi.org/10.1002/nme.7473}

\bibitem{Phan2025_5f41}
N.~N. Phan, W.~Sun, J.~D. Clayton, \href{https://doi.org/10.1016/j.cma.2025.117792}{Hydra: Symbolic feature engineering of overparameterized eulerian hyperelasticity models for fast inference time}, Computer Methods in Applied Mechanics and Engineering 437 (2025) 117792.
\newblock \href {https://doi.org/10.1016/j.cma.2025.117792} {\path{doi:10.1016/j.cma.2025.117792}}.
\newline\urlprefix\url{https://doi.org/10.1016/j.cma.2025.117792}

\bibitem{Sun2025_13f9}
W.~Sun, N.~N. Phan, \href{https://doi.org/10.1007/978-3-031-93213-7_39}{Discovery of Symbolic Hyperelasticity Models for Anisotropic Solids Beyond Linear Combinations}, Springer Nature Switzerland, 2025, pp. 497--510.
\newline\urlprefix\url{https://doi.org/10.1007/978-3-031-93213-7_39}

\bibitem{Linka2021_1735}
K.~Linka, M.~Hillg{\"a}rtner, K.~P. Abdolazizi, R.~C. Aydin, M.~Itskov, C.~J. Cyron, \href{http://dx.doi.org/10.1016/j.jcp.2020.110010}{Constitutive artificial neural networks: A fast and general approach to predictive data-driven constitutive modeling by deep learning}, Journal of Computational Physics 429 (2021) 110010.
\newblock \href {https://doi.org/10.1016/j.jcp.2020.110010} {\path{doi:10.1016/j.jcp.2020.110010}}.
\newline\urlprefix\url{http://dx.doi.org/10.1016/j.jcp.2020.110010}

\bibitem{Linka2023_6533}
K.~Linka, E.~Kuhl, \href{https://doi.org/10.1016/j.cma.2022.115731}{A new family of constitutive artificial neural networks towards automated model discovery}, Computer Methods in Applied Mechanics and Engineering 403 (2023) 115731.
\newblock \href {https://doi.org/10.1016/j.cma.2022.115731} {\path{doi:10.1016/j.cma.2022.115731}}.
\newline\urlprefix\url{https://doi.org/10.1016/j.cma.2022.115731}

\bibitem{Linka2023_1ca2}
K.~Linka, S.~R. St.~Pierre, E.~Kuhl, \href{https://doi.org/10.1016/j.actbio.2023.01.055}{Automated model discovery for human brain using constitutive artificial neural networks}, Acta Biomaterialia 160 (2023) 134--151.
\newblock \href {https://doi.org/10.1016/j.actbio.2023.01.055} {\path{doi:10.1016/j.actbio.2023.01.055}}.
\newline\urlprefix\url{https://doi.org/10.1016/j.actbio.2023.01.055}

\bibitem{Peirlinck2024_f38b}
M.~Peirlinck, K.~Linka, J.~A. Hurtado, E.~Kuhl, \href{https://doi.org/10.1016/j.cma.2023.116534}{On automated model discovery and a universal material subroutine for hyperelastic materials}, Computer Methods in Applied Mechanics and Engineering 418 (2024) 116534.
\newblock \href {https://doi.org/10.1016/j.cma.2023.116534} {\path{doi:10.1016/j.cma.2023.116534}}.
\newline\urlprefix\url{https://doi.org/10.1016/j.cma.2023.116534}

\bibitem{St_Pierre2023}
S.~R. St.~Pierre, K.~Linka, E.~Kuhl, \href{https://doi.org/10.1016/j.brain.2023.100066}{Principal-stretch-based constitutive neural networks autonomously discover a subclass of ogden models for human brain tissue}, Brain Multiphysics 4 (2023) 100066.
\newblock \href {https://doi.org/10.1016/j.brain.2023.100066} {\path{doi:10.1016/j.brain.2023.100066}}.
\newline\urlprefix\url{https://doi.org/10.1016/j.brain.2023.100066}

\bibitem{Linka2023_1bb2}
K.~Linka, A.~Buganza~Tepole, G.~A. Holzapfel, E.~Kuhl, \href{https://doi.org/10.1016/j.cma.2023.116007}{Automated model discovery for skin: Discovering the best model, data, and experiment}, Computer Methods in Applied Mechanics and Engineering 410 (2023) 116007.
\newblock \href {https://doi.org/10.1016/j.cma.2023.116007} {\path{doi:10.1016/j.cma.2023.116007}}.
\newline\urlprefix\url{https://doi.org/10.1016/j.cma.2023.116007}

\bibitem{St_Pierre2023_1c60}
S.~R. St.~Pierre, D.~Rajasekharan, E.~C. Darwin, K.~Linka, M.~E. Levenston, E.~Kuhl, \href{https://doi.org/10.1016/j.cma.2023.116236}{Discovering the mechanics of artificial and real meat}, Computer Methods in Applied Mechanics and Engineering 415 (2023) 116236.
\newblock \href {https://doi.org/10.1016/j.cma.2023.116236} {\path{doi:10.1016/j.cma.2023.116236}}.
\newline\urlprefix\url{https://doi.org/10.1016/j.cma.2023.116236}

\bibitem{Peirlinck2024_34b9}
M.~Peirlinck, K.~Linka, J.~A. Hurtado, G.~A. Holzapfel, E.~Kuhl, Democratizing biomedical simulation through automated model discovery and a universal material subroutine, Computational mechanics (2024) 1--21.

\bibitem{Vervenne2025_31ed}
T.~Vervenne, M.~Peirlinck, N.~Famaey, E.~Kuhl, \href{https://doi.org/10.1007/s10237-025-01930-1}{Constitutive neural networks for main pulmonary arteries: discovering the undiscovered}, Biomechanics and Modeling in Mechanobiology 24 (2025) 615--634.
\newblock \href {https://doi.org/10.1007/s10237-025-01930-1} {\path{doi:10.1007/s10237-025-01930-1}}.
\newline\urlprefix\url{https://doi.org/10.1007/s10237-025-01930-1}

\bibitem{Martonova2024_9709}
D.~Martonov{\'a}, M.~Peirlinck, K.~Linka, G.~A. Holzapfel, S.~Leyendecker, E.~Kuhl, \href{https://doi.org/10.1016/j.cma.2024.117078}{Automated model discovery for human cardiac tissue: Discovering the best model and parameters}, Computer Methods in Applied Mechanics and Engineering 428 (2024) 117078.
\newblock \href {https://doi.org/10.1016/j.cma.2024.117078} {\path{doi:10.1016/j.cma.2024.117078}}.
\newline\urlprefix\url{https://doi.org/10.1016/j.cma.2024.117078}

\bibitem{Peirlinck2025_53e6}
M.~Peirlinck, K.~Linka, E.~Kuhl, \href{http://arxiv.org/abs/2504.02748}{Atrial constitutive neural networks} (4 2025).
\newblock \href {http://arxiv.org/abs/2504.02748} {\path{arXiv:2504.02748}}, \href {https://doi.org/10.48550/arXiv.2504.02748} {\path{doi:10.48550/arXiv.2504.02748}}.
\newline\urlprefix\url{http://arxiv.org/abs/2504.02748}

\bibitem{Wang2023_4e9b}
L.~M. Wang, K.~Linka, E.~Kuhl, \href{https://doi.org/10.1016/j.jmbbm.2023.106021}{Automated model discovery for muscle using constitutive recurrent neural networks}, Journal of the Mechanical Behavior of Biomedical Materials 145 (2023) 106021.
\newblock \href {https://doi.org/10.1016/j.jmbbm.2023.106021} {\path{doi:10.1016/j.jmbbm.2023.106021}}.
\newline\urlprefix\url{https://doi.org/10.1016/j.jmbbm.2023.106021}

\bibitem{Abdolazizi2024_4363}
K.~P. Abdolazizi, K.~Linka, C.~J. Cyron, \href{https://doi.org/10.1016/j.jcp.2023.112704}{Viscoelastic constitutive artificial neural networks (vcanns) \textendash a framework for data-driven anisotropic nonlinear finite viscoelasticity}, Journal of Computational Physics 499 (2024) 112704.
\newblock \href {https://doi.org/10.1016/j.jcp.2023.112704} {\path{doi:10.1016/j.jcp.2023.112704}}.
\newline\urlprefix\url{https://doi.org/10.1016/j.jcp.2023.112704}

\bibitem{Holthusen2025_142a}
H.~Holthusen, T.~Brepols, K.~Linka, E.~Kuhl, \href{https://doi.org/10.1016/j.compbiomed.2025.109691}{Automated model discovery for tensional homeostasis: Constitutive machine learning in growth and remodeling}, Computers in Biology and Medicine 186 (2025) 109691.
\newblock \href {https://doi.org/10.1016/j.compbiomed.2025.109691} {\path{doi:10.1016/j.compbiomed.2025.109691}}.
\newline\urlprefix\url{https://doi.org/10.1016/j.compbiomed.2025.109691}

\bibitem{Linka2025_32dc}
K.~Linka, G.~A. Holzapfel, E.~Kuhl, \href{https://doi.org/10.1016/j.cma.2024.117517}{Discovering uncertainty: Bayesian constitutive artificial neural networks}, Computer Methods in Applied Mechanics and Engineering 433 (2025) 117517.
\newblock \href {https://doi.org/10.1016/j.cma.2024.117517} {\path{doi:10.1016/j.cma.2024.117517}}.
\newline\urlprefix\url{https://doi.org/10.1016/j.cma.2024.117517}

\bibitem{Peirlinck2024_55e5}
M.~Peirlinck, J.~A. Hurtado, M.~K. Rausch, A.~B. Tepole, E.~Kuhl, \href{https://doi.org/10.1007/s00366-024-02031-w}{A universal material model subroutine for soft matter systems}, Engineering with Computers (9 2024).
\newblock \href {https://doi.org/10.1007/s00366-024-02031-w} {\path{doi:10.1007/s00366-024-02031-w}}.
\newline\urlprefix\url{https://doi.org/10.1007/s00366-024-02031-w}

\bibitem{Holzapfel2000_7a2d}
G.~A. Holzapfel, T.~C. Gasser, R.~W. Ogden, \href{https://doi.org/10.1023/a:1010835316564}{A new constitutive framework for arterial wall mechanics and a comparative study of material models}, Journal of Elasticity 61 (2000) 1--48.
\newblock \href {https://doi.org/10.1023/a:1010835316564} {\path{doi:10.1023/a:1010835316564}}.
\newline\urlprefix\url{https://doi.org/10.1023/a:1010835316564}

\bibitem{Gasser2006_3595}
T.~C. Gasser, R.~W. Ogden, G.~A. Holzapfel, \href{http://dx.doi.org/10.1098/rsif.2005.0073}{Hyperelastic modelling of arterial layers with distributed collagen fibre orientations}, Journal of The Royal Society Interface 3 (2006) 15--35.
\newblock \href {https://doi.org/10.1098/rsif.2005.0073} {\path{doi:10.1098/rsif.2005.0073}}.
\newline\urlprefix\url{http://dx.doi.org/10.1098/rsif.2005.0073}

\bibitem{Holzapfel2009_4559}
G.~A. Holzapfel, R.~W. Ogden, \href{https://doi.org/10.1098/rsta.2009.0091}{Constitutive modelling of passive myocardium: a structurally based framework for material characterization}, Philosophical Transactions of the Royal Society A: Mathematical, Physical and Engineering Sciences 367 (2009) 3445--3475.
\newblock \href {https://doi.org/10.1098/rsta.2009.0091} {\path{doi:10.1098/rsta.2009.0091}}.
\newline\urlprefix\url{https://doi.org/10.1098/rsta.2009.0091}

\bibitem{Avazmohammadi2019}
R.~Avazmohammadi, J.~S. Soares, D.~S. Li, S.~S. Raut, R.~C. Gorman, M.~S. Sacks, \href{https://doi.org/10.1146/annurev-bioeng-062117-121129}{A contemporary look at biomechanical models of myocardium}, Annual Review of Biomedical Engineering 21 (2019) 417--442.
\newblock \href {https://doi.org/10.1146/annurev-bioeng-062117-121129} {\path{doi:10.1146/annurev-bioeng-062117-121129}}.
\newline\urlprefix\url{https://doi.org/10.1146/annurev-bioeng-062117-121129}

\bibitem{Aggarwal2023}
A.~Aggarwal, L.~T. Hudson, D.~W. Laurence, C.-H. Lee, S.~Pant, \href{http://dx.doi.org/10.1016/j.jmbbm.2023.105657}{A bayesian constitutive model selection framework for biaxial mechanical testing of planar soft tissues: Application to porcine aortic valves}, Journal of the Mechanical Behavior of Biomedical Materials 138 (2023) 105657.
\newblock \href {https://doi.org/10.1016/j.jmbbm.2023.105657} {\path{doi:10.1016/j.jmbbm.2023.105657}}.
\newline\urlprefix\url{http://dx.doi.org/10.1016/j.jmbbm.2023.105657}

\bibitem{Krijnen2026_52b3}
R.~P. Krijnen, A.~Joshi, S.~Kumar, M.~Peirlinck, \href{https://doi.org/10.1016/j.cma.2026.119034}{Unsupervised full-field bayesian inference of orthotropic hyperelasticity from a single biaxial test: a myocardial case study}, Computer Methods in Applied Mechanics and Engineering 459 (2026) 119034.
\newblock \href {https://doi.org/10.1016/j.cma.2026.119034} {\path{doi:10.1016/j.cma.2026.119034}}.
\newline\urlprefix\url{https://doi.org/10.1016/j.cma.2026.119034}

\bibitem{Famaey2026_7035}
N.~Famaey, H.~Fehervary, Y.~Lafon, A.~Akyildiz, S.~Dreesen, K.~Bruy{\`e}re-Garnier, J.-M. Allain, M.~Alloisio, A.~Aparici-Gil, C.~Catalano, F.~Chassagne, S.~Chokhandre, K.~Crevits, H.~Crielaard, E.~Cunnane, C.~Cunnane, K.~De~Leener, A.~Desai, R.~Driessen, A.~Erdemir, M.~Eskandari, S.~Evans, C.~Gasser, M.~Gebhardt, B.~Glasmacher, G.~A. Holzapfel, M.~Isasi, L.~Jennings, S.~Kurz, S.~Leal-Marin, P.~Lecomte, A.~Morch, J.~Mulvihill, F.~Nemavhola, T.~Pandelani, S.~Pasta, E.~Pe{\~n}a, B.~Pierrat, H.-L. Ploeg, S.~Polzer, M.~Rausch, D.~Schwarz, H.~Screen, S.~Sherifova, G.~Sommer, S.~Wang, D.~Walsh, D.~Yadav, T.~Marchal, L.~Geris, \href{https://doi.org/10.1016/j.jbiomech.2025.113021}{Community challenge towards consensus on characterization of biological tissue: C4bio’s first findings}, Journal of Biomechanics 194 (2026) 113021.
\newblock \href {https://doi.org/10.1016/j.jbiomech.2025.113021} {\path{doi:10.1016/j.jbiomech.2025.113021}}.
\newline\urlprefix\url{https://doi.org/10.1016/j.jbiomech.2025.113021}

\bibitem{Pierron2012_54c0}
F.~Pierron, M.~Gr{\'e}diac, The virtual fields method: extracting constitutive mechanical parameters from full-field deformation measurements, Springer Science \textbackslash \& Business Media, 2012.

\bibitem{Kolawole2023}
F.~O. Kolawole, M.~Peirlinck, T.~E. Cork, M.~Levenston, E.~Kuhl, D.~B. Ennis, \href{http://dx.doi.org/10.1007/s10439-023-03164-7}{Validating mri-derived myocardial stiffness estimates using in vitro synthetic heart models}, Annals of Biomedical Engineering 51~(7) (2023) 1574–1587.
\newblock \href {https://doi.org/10.1007/s10439-023-03164-7} {\path{doi:10.1007/s10439-023-03164-7}}.
\newline\urlprefix\url{http://dx.doi.org/10.1007/s10439-023-03164-7}

\bibitem{Meng2025_1d5e}
S.~Meng, A.~A.~K. Yousefi, S.~Avril, \href{https://doi.org/10.1016/j.cma.2024.117580}{Machine-learning-based virtual fields method: Application to anisotropic hyperelasticity}, Computer Methods in Applied Mechanics and Engineering 434 (2025) 117580.
\newblock \href {https://doi.org/10.1016/j.cma.2024.117580} {\path{doi:10.1016/j.cma.2024.117580}}.
\newline\urlprefix\url{https://doi.org/10.1016/j.cma.2024.117580}

\bibitem{Navy2025_7f0d}
X.~Navy, Z.~Sheng, K.~Kim, J.~M. Cormack, \href{https://doi.org/10.1016/j.ultrasmedbio.2025.05.007}{Three-dimensional tissue strain measurement using a row\textendash column array during biaxial testing of excised ventricular porcine myocardium}, Ultrasound in Medicine \&amp; Biology 51 (2025) 1622--1626.
\newblock \href {https://doi.org/10.1016/j.ultrasmedbio.2025.05.007} {\path{doi:10.1016/j.ultrasmedbio.2025.05.007}}.
\newline\urlprefix\url{https://doi.org/10.1016/j.ultrasmedbio.2025.05.007}

\bibitem{Kumar2025_5768}
S.~Kumar, D.~T. Seidl, B.~N. Granzow, J.~Yang, J.~N. Fuhg, \href{https://doi.org/10.1016/j.cma.2025.118159}{A comparative study of calibration techniques for finite strain elastoplasticity: Numerically-exact sensitivities for femu and vfm}, Computer Methods in Applied Mechanics and Engineering 444 (2025) 118159.
\newblock \href {https://doi.org/10.1016/j.cma.2025.118159} {\path{doi:10.1016/j.cma.2025.118159}}.
\newline\urlprefix\url{https://doi.org/10.1016/j.cma.2025.118159}

\bibitem{Knipper2026_560d}
M.~Knipper, C.~Ji, M.~Brand, K.~Linka, \href{http://arxiv.org/abs/2606.05199}{Finite element-based material learning via automatic differentiation: Learning constitutive neural network models from full-field deformation data} (5 2026).
\newblock \href {http://arxiv.org/abs/2606.05199} {\path{arXiv:2606.05199}}, \href {https://doi.org/10.48550/arXiv.2606.05199} {\path{doi:10.48550/arXiv.2606.05199}}.
\newline\urlprefix\url{http://arxiv.org/abs/2606.05199}

\bibitem{grediac1989principe}
M.~Gr{\'e}diac, Principe des travaux virtuels et identification, Comptes rendus de l\textquotesingle Acad{\textbackslash \textquotesingle e}mie dessciences. S{\textbackslash \textquotesingle e}rie 2, M{\textbackslash \textquotesingle e}canique, Physique, Chimie, Sciences de l\textquotesingle univers, Sciences de la Terre 309~(1) (1989) 1--5.

\bibitem{Grediac2006_592b}
M.~Gr{\'e}diac, F.~Pierron, S.~Avril, E.~Toussaint, \href{https://doi.org/10.1111/j.1475-1305.2006.tb01504.x}{The virtual fields method for extracting constitutive parameters from full-field measurements: a review}, Strain 42 (2006) 233--253.
\newblock \href {https://doi.org/10.1111/j.1475-1305.2006.tb01504.x} {\path{doi:10.1111/j.1475-1305.2006.tb01504.x}}.
\newline\urlprefix\url{https://doi.org/10.1111/j.1475-1305.2006.tb01504.x}

\bibitem{Avril2026_1028}
S.~Avril, \href{https://doi.org/10.1115/1.4071211}{Recent advances in the virtual fields method for evaluating and identifying tissue biomechanical properties and constitutive parameters}, Journal of Biomechanical Engineering 148 (5 2026).
\newblock \href {https://doi.org/10.1115/1.4071211} {\path{doi:10.1115/1.4071211}}.
\newline\urlprefix\url{https://doi.org/10.1115/1.4071211}

\bibitem{Claire2004_7994}
D.~Claire, F.~Hild, S.~Roux, \href{https://doi.org/10.1002/nme.1057}{A finite element formulation to identify damage fields: the equilibrium gap method}, International Journal for Numerical Methods in Engineering 61 (2004) 189--208.
\newblock \href {https://doi.org/10.1002/nme.1057} {\path{doi:10.1002/nme.1057}}.
\newline\urlprefix\url{https://doi.org/10.1002/nme.1057}

\bibitem{Amiot2013_290b}
F.~Amiot, J.~P{\'e}ri{\'e}, S.~Roux, \href{https://doi.org/10.1002/9781118578469.ch12}{Equilibrium gap method} (4 2013).
\newblock \href {https://doi.org/10.1002/9781118578469.ch12} {\path{doi:10.1002/9781118578469.ch12}}.
\newline\urlprefix\url{https://doi.org/10.1002/9781118578469.ch12}

\bibitem{Flaschel2021_4e47}
M.~Flaschel, S.~Kumar, L.~De~Lorenzis, \href{http://dx.doi.org/10.1016/j.cma.2021.113852}{Unsupervised discovery of interpretable hyperelastic constitutive laws}, Computer Methods in Applied Mechanics and Engineering 381 (2021) 113852.
\newblock \href {https://doi.org/10.1016/j.cma.2021.113852} {\path{doi:10.1016/j.cma.2021.113852}}.
\newline\urlprefix\url{http://dx.doi.org/10.1016/j.cma.2021.113852}

\bibitem{Demiray1972}
H.~Demiray, \href{http://dx.doi.org/10.1016/0021-9290(72)90047-4}{A note on the elasticity of soft biological tissues}, Journal of Biomechanics 5~(3) (1972) 309–311.
\newblock \href {https://doi.org/10.1016/0021-9290(72)90047-4} {\path{doi:10.1016/0021-9290(72)90047-4}}.
\newline\urlprefix\url{http://dx.doi.org/10.1016/0021-9290(72)90047-4}

\bibitem{Humphrey1987_2c2a}
J.~D. Humphrey, F.~C.~P. Yin, \href{https://doi.org/10.1115/1.3138684}{On constitutive relations and finite deformations of passive cardiac tissue: I. a pseudostrain-energy function}, Journal of Biomechanical Engineering 109 (1987) 298--304.
\newblock \href {https://doi.org/10.1115/1.3138684} {\path{doi:10.1115/1.3138684}}.
\newline\urlprefix\url{https://doi.org/10.1115/1.3138684}

\bibitem{Guccione1991_138b}
J.~M. Guccione, A.~D. McCulloch, L.~K. Waldman, \href{https://doi.org/10.1115/1.2894084}{Passive material properties of intact ventricular myocardium determined from a cylindrical model}, Journal of Biomechanical Engineering 113 (1991) 42--55.
\newblock \href {https://doi.org/10.1115/1.2894084} {\path{doi:10.1115/1.2894084}}.
\newline\urlprefix\url{https://doi.org/10.1115/1.2894084}

\bibitem{Holzapfel2000_73bc}
G.~A. Holzapfel, Nonlinear solid mechanics: a continuum approach for engineering, Wiley, 2000.

\bibitem{Thakolkaran2022_37f6}
P.~Thakolkaran, A.~Joshi, Y.~Zheng, M.~Flaschel, L.~De~Lorenzis, S.~Kumar, \href{http://dx.doi.org/10.1016/j.jmps.2022.105076}{Nn-euclid: Deep-learning hyperelasticity without stress data}, Journal of the Mechanics and Physics of Solids 169 (2022) 105076.
\newblock \href {https://doi.org/10.1016/j.jmps.2022.105076} {\path{doi:10.1016/j.jmps.2022.105076}}.
\newline\urlprefix\url{http://dx.doi.org/10.1016/j.jmps.2022.105076}

\bibitem{Flaschel2022_7c13}
M.~Flaschel, S.~Kumar, L.~De~Lorenzis, \href{http://dx.doi.org/10.1038/s41524-022-00752-4}{Discovering plasticity models without stress data}, npj Computational Materials 8 (2022) 1--10.
\newblock \href {https://doi.org/10.1038/s41524-022-00752-4} {\path{doi:10.1038/s41524-022-00752-4}}.
\newline\urlprefix\url{http://dx.doi.org/10.1038/s41524-022-00752-4}

\bibitem{Xu2025_4533}
H.~Xu, M.~Flaschel, L.~De~Lorenzis, \href{https://doi.org/10.1186/s40323-024-00281-3}{Discovering non-associated pressure-sensitive plasticity models with euclid}, Advanced Modeling and Simulation in Engineering Sciences 12 (2025) 1.
\newblock \href {https://doi.org/10.1186/s40323-024-00281-3} {\path{doi:10.1186/s40323-024-00281-3}}.
\newline\urlprefix\url{https://doi.org/10.1186/s40323-024-00281-3}

\bibitem{Marino2023_5b43}
E.~Marino, M.~Flaschel, S.~Kumar, L.~De~Lorenzis, \href{http://dx.doi.org/10.1016/j.mechmat.2023.104643}{Automated identification of linear viscoelastic constitutive laws with euclid}, Mechanics of Materials 181 (2023) 104643.
\newblock \href {https://doi.org/10.1016/j.mechmat.2023.104643} {\path{doi:10.1016/j.mechmat.2023.104643}}.
\newline\urlprefix\url{http://dx.doi.org/10.1016/j.mechmat.2023.104643}

\bibitem{Flaschel2023_564d}
M.~Flaschel, S.~Kumar, L.~De~Lorenzis, \href{https://www.sciencedirect.com/science/article/pii/S0045782522008234}{Automated discovery of generalized standard material models with euclid}, Computer Methods in Applied Mechanics and Engineering 405 (2023) 115867.
\newblock \href {https://doi.org/10.1016/j.cma.2022.115867} {\path{doi:10.1016/j.cma.2022.115867}}.
\newline\urlprefix\url{https://www.sciencedirect.com/science/article/pii/S0045782522008234}

\bibitem{Joshi2022_335f}
A.~Joshi, P.~Thakolkaran, Y.~Zheng, M.~Escande, M.~Flaschel, L.~De~Lorenzis, S.~Kumar, \href{http://dx.doi.org/10.1016/j.cma.2022.115225}{Bayesian-euclid: Discovering hyperelastic material laws with uncertainties}, Computer Methods in Applied Mechanics and Engineering 398 (2022) 115225.
\newblock \href {https://doi.org/10.1016/j.cma.2022.115225} {\path{doi:10.1016/j.cma.2022.115225}}.
\newline\urlprefix\url{http://dx.doi.org/10.1016/j.cma.2022.115225}

\bibitem{Martonova2025_66e5}
D.~Martonov{\'a}, S.~Leyendecker, G.~A. Holzapfel, E.~Kuhl, \href{https://doi.org/10.1007/s10237-025-02005-x}{Discovering dispersion: How robust is automated model discovery for human myocardial tissue?}, Biomechanics and Modeling in Mechanobiology 24 (2025) 2023--2037.
\newblock \href {https://doi.org/10.1007/s10237-025-02005-x} {\path{doi:10.1007/s10237-025-02005-x}}.
\newline\urlprefix\url{https://doi.org/10.1007/s10237-025-02005-x}

\bibitem{Martonova2026_7570}
D.~Martonov{\'a}, A.~Goriely, E.~Kuhl, \href{https://doi.org/10.1016/j.jmps.2025.106352}{Generalized invariants meet constitutive neural networks: A novel framework for hyperelastic materials}, Journal of the Mechanics and Physics of Solids 206 (2026) 106352.
\newblock \href {https://doi.org/10.1016/j.jmps.2025.106352} {\path{doi:10.1016/j.jmps.2025.106352}}.
\newline\urlprefix\url{https://doi.org/10.1016/j.jmps.2025.106352}

\bibitem{Ball1976_4d15}
J.~M. Ball, \href{https://doi.org/10.1007/bf00279992}{Convexity conditions and existence theorems in nonlinear elasticity}, Archive for Rational Mechanics and Analysis 63 (1976) 337--403.
\newblock \href {https://doi.org/10.1007/bf00279992} {\path{doi:10.1007/bf00279992}}.
\newline\urlprefix\url{https://doi.org/10.1007/bf00279992}

\bibitem{Martonova2025}
D.~Martonov{\'a}, A.~Goriely, E.~Kuhl, Generalized invariants meet constitutive neural networks: A novel framework for hyperelastic materials (8 2025).

\bibitem{Schroder2003_1753}
J.~Schr{\"o}der, P.~Neff, \href{http://dx.doi.org/10.1016/s0020-7683(02)00458-4}{Invariant formulation of hyperelastic transverse isotropy based on polyconvex free energy functions}, International Journal of Solids and Structures 40 (2003) 401--445.
\newblock \href {https://doi.org/10.1016/s0020-7683(02)00458-4} {\path{doi:10.1016/s0020-7683(02)00458-4}}.
\newline\urlprefix\url{http://dx.doi.org/10.1016/s0020-7683(02)00458-4}

\bibitem{Balzani2006_50ce}
D.~Balzani, P.~Neff, J.~Schr{\"o}der, G.~Holzapfel, \href{http://dx.doi.org/10.1016/j.ijsolstr.2005.07.048}{A polyconvex framework for soft biological tissues. adjustment to experimental data}, International Journal of Solids and Structures 43 (2006) 6052--6070.
\newblock \href {https://doi.org/10.1016/j.ijsolstr.2005.07.048} {\path{doi:10.1016/j.ijsolstr.2005.07.048}}.
\newline\urlprefix\url{http://dx.doi.org/10.1016/j.ijsolstr.2005.07.048}

\bibitem{Frank1993_6411}
l.~E. Frank, J.~H. Friedman, \href{https://doi.org/10.1080/00401706.1993.10485033}{A statistical view of some chemometrics regression tools}, Technometrics 35 (1993) 109--135.
\newblock \href {https://doi.org/10.1080/00401706.1993.10485033} {\path{doi:10.1080/00401706.1993.10485033}}.
\newline\urlprefix\url{https://doi.org/10.1080/00401706.1993.10485033}

\bibitem{Hoerl1970_614f}
A.~E. Hoerl, R.~W. Kennard, \href{https://doi.org/10.1080/00401706.1970.10488634}{Ridge regression: Biased estimation for nonorthogonal problems}, Technometrics 12 (1970) 55--67.
\newblock \href {https://doi.org/10.1080/00401706.1970.10488634} {\path{doi:10.1080/00401706.1970.10488634}}.
\newline\urlprefix\url{https://doi.org/10.1080/00401706.1970.10488634}

\bibitem{McCulloch2024_6c1e}
J.~A. McCulloch, S.~R. St.~Pierre, K.~Linka, E.~Kuhl, On sparse regression, lp-regularization, and automated model discovery, International Journal for Numerical Methods in Engineering 125~(14) (2024) e7481.

\bibitem{Rivlin1948_7623}
R.~S. Rivlin, \href{https://doi.org/10.1098/rsta.1948.0024}{Large elastic deformations of isotropic materials iv. further developments of the general theory}, Philosophical Transactions of the Royal Society of London. Series A, Mathematical and Physical Sciences 241 (1948) 379--397.
\newblock \href {https://doi.org/10.1098/rsta.1948.0024} {\path{doi:10.1098/rsta.1948.0024}}.
\newline\urlprefix\url{https://doi.org/10.1098/rsta.1948.0024}

\bibitem{Isihara1951_5512}
A.~Isihara, N.~Hashitsume, M.~Tatibana, \href{https://doi.org/10.1063/1.1748111}{Statistical theory of rubber‐like elasticity. iv. (two‐dimensional stretching)}, The Journal of Chemical Physics 19 (1951) 1508--1512.
\newblock \href {https://doi.org/10.1063/1.1748111} {\path{doi:10.1063/1.1748111}}.
\newline\urlprefix\url{https://doi.org/10.1063/1.1748111}

\bibitem{Arruda1993_3631}
E.~M. Arruda, M.~C. Boyce, \href{https://doi.org/10.1016/0022-5096(93)90013-6}{A three-dimensional constitutive model for the large stretch behavior of rubber elastic materials}, Journal of the Mechanics and Physics of Solids 41 (1993) 389--412.
\newblock \href {https://doi.org/10.1016/0022-5096(93)90013-6} {\path{doi:10.1016/0022-5096(93)90013-6}}.
\newline\urlprefix\url{https://doi.org/10.1016/0022-5096(93)90013-6}

\bibitem{Gent1958_2fcf}
A.~N. Gent, A.~G. Thomas, \href{https://doi.org/10.1002/pol.1958.1202811814}{Forms for the stored (strain) energy function for vulcanized rubber}, Journal of Polymer Science 28 (1958) 625--628.
\newblock \href {https://doi.org/10.1002/pol.1958.1202811814} {\path{doi:10.1002/pol.1958.1202811814}}.
\newline\urlprefix\url{https://doi.org/10.1002/pol.1958.1202811814}

\bibitem{Ogden1972_142d}
R.~W. Ogden, \href{https://doi.org/10.1098/rspa.1972.0026}{Large deformation isotropic elasticity \textendash on the correlation of theory and experiment for incompressible rubberlike solids}, Proceedings of the Royal Society of London. A. Mathematical and Physical Sciences 326 (1972) 565--584.
\newblock \href {https://doi.org/10.1098/rspa.1972.0026} {\path{doi:10.1098/rspa.1972.0026}}.
\newline\urlprefix\url{https://doi.org/10.1098/rspa.1972.0026}

\bibitem{Ning2006_6dc7}
X.~Ning, Q.~Zhu, Y.~Lanir, S.~S. Margulies, \href{https://doi.org/10.1115/1.2354208}{A transversely isotropic viscoelastic constitutive equation for brainstem undergoing finite deformation}, Journal of Biomechanical Engineering 128 (2006) 925--933.
\newblock \href {https://doi.org/10.1115/1.2354208} {\path{doi:10.1115/1.2354208}}.
\newline\urlprefix\url{https://doi.org/10.1115/1.2354208}

\bibitem{Meaney2003_534c}
D.~F. Meaney, \href{https://doi.org/10.1007/s10237-002-0020-1}{Relationship between structural modeling and hyperelastic material behavior: application to cns white matter}, Biomechanics and Modeling in Mechanobiology 1 (2003) 279--293.
\newblock \href {https://doi.org/10.1007/s10237-002-0020-1} {\path{doi:10.1007/s10237-002-0020-1}}.
\newline\urlprefix\url{https://doi.org/10.1007/s10237-002-0020-1}

\bibitem{Merodio2005_4c0c}
J.~Merodio, R.~Ogden, \href{https://doi.org/10.1016/j.ijnonlinmec.2004.05.003}{Mechanical response of fiber-reinforced incompressible non-linearly elastic solids}, International Journal of Non-Linear Mechanics 40 (2005) 213--227.
\newblock \href {https://doi.org/10.1016/j.ijnonlinmec.2004.05.003} {\path{doi:10.1016/j.ijnonlinmec.2004.05.003}}.
\newline\urlprefix\url{https://doi.org/10.1016/j.ijnonlinmec.2004.05.003}

\bibitem{Amos2017_773c}
B.~Amos, L.~Xu, J.~Z. Kolter, \href{https://proceedings.mlr.press/v70/amos17b.html}{Input convex neural networks}, in: Proceedings of the 34th International Conference on Machine Learning, PMLR, 2017, pp. 146--155.
\newline\urlprefix\url{https://proceedings.mlr.press/v70/amos17b.html}

\bibitem{morrey1952quasi}
J.~Morrey, B.~Charles, Quasi-convexity and the lower semicontinuity of multiple integrals, Pacific Journal of Mathematics 2 (1952) 25–53.

\bibitem{As_ad2022_6b13}
F.~As\textquotesingle~ad, P.~Avery, C.~Farhat, \href{http://dx.doi.org/10.1002/nme.6957}{A mechanics-informed artificial neural network approach in data-driven constitutive modeling}, International Journal for Numerical Methods in Engineering 123 (2022) 2738--2759.
\newblock \href {https://doi.org/10.1002/nme.6957} {\path{doi:10.1002/nme.6957}}.
\newline\urlprefix\url{http://dx.doi.org/10.1002/nme.6957}

\bibitem{LeCun1998_77c8}
Y.~LeCun, L.~Bottou, G.~B. Orr, K.~R. M{\"u}ller, \href{https://doi.org/10.1007/3-540-49430-8_2}{Neural Networks: Tricks of the Trade}, Springer, 1998, Ch. Efficient BackProp, pp. 9--50.
\newline\urlprefix\url{https://doi.org/10.1007/3-540-49430-8_2}

\bibitem{Huang2023_681f}
L.~Huang, J.~Qin, Y.~Zhou, F.~Zhu, L.~Liu, L.~Shao, \href{https://doi.org/10.1109/tpami.2023.3250241}{Normalization techniques in training dnns: Methodology, analysis and application}, IEEE Transactions on Pattern Analysis and Machine Intelligence 45 (2023) 10173--10196.
\newblock \href {https://doi.org/10.1109/tpami.2023.3250241} {\path{doi:10.1109/tpami.2023.3250241}}.
\newline\urlprefix\url{https://doi.org/10.1109/tpami.2023.3250241}

\end{thebibliography}

\end{document}